# Anion vacancy driven magnetism in incipient ferroelectric SrTiO$_3$ and KTaO$_3$ nanoparticles

*E.A. Eliseev*[1], *A. N. Morozovska*[1,2*], *and M.D. Glinchuk*[1†], *R. Blinc*[3]

[1] Institute for Problems of Materials Science, National Academy of Sciences of Ukraine,
Krjijanovskogo 3, 03142 Kiev, Ukraine,

[2] V. Lashkarev Institute of Semiconductor Physics, National Academy of Sciences of Ukraine,
prospect Nauki 41, 03028 Kiev, Ukraine

[3] Jožef Stefan Institute, P. O. Box 3000, 1001 Ljubljana, Slovenia

*Abstract*

Based on our analytical results [http://arxiv.org/abs/1006.3670], we predict that undoped nanoparticles (size ≤ 10 − 100 nm) of incipient ferroelectrics without any magnetic ions can become ferromagnetic even at room temperatures due to the inherent presence of a **new type of magnetic defects** with spin *S*=1, namely oxygen vacancies, where the magnetic triplet state is the ground state in the vicinity of the surface (**magnetic shell**), while the nonmagnetic singlet is the ground state in the bulk material (**nonmagnetic core**).

Consideration of randomly distributed magnetic spins (S=1) had shown that magnetic properties of incipient ferroelectric nanoparticles are strongly size and temperature dependent due to the size and temperature dependence of their dielectric permittivity $\varepsilon(T,R)$ and the effective Bohr radius $a_B^*(T,R) \sim \varepsilon(T,R)$. The phase diagrams in coordinates temperature and particle radius are considered. In particular, for particle radii *R* less that the critical radius $R_c(T)$ ferromagnetic long-range order appears in a shell region of thickness 5 – 50 nm once the concentration of magnetic defects *N* exceeds the magnetic percolation threshold $N^p$. The critical radius $R_c(T)$ is calculated in the mean field theory from the condition $J(r, R_c, T) = k_B T$, where $J(r, R)$ is the exchange energy of the magnetic defects, and *r* is the average distance between the defects. The strong size and temperature dependence of the exchange energy originates from the dependence $\varepsilon(T,R)$. At vacancy concentrations $N < N^p$ and radii $R < R_c(T)$ short-range ferromagnetic order and consequently a glass state may appear. For particle radii $R > R_c(T)$ only the paramagnetic phase is possible. The conditions of the super-paramagnetic state appearance in the assembly of nanoparticles with narrow distribution function of their sizes are discussed also.

---

[*] Corresponding author: morozo@i.com.ua
[†] Corresponding author: glin@ipms.kiev.ua

*1. Introduction*

Numerous experiments revealed ferromagnetic properties of nanomaterials, which are nonmagnetic in the bulk. For instance, remarkable room-temperature ferromagnetism was observed in undoped $TiO_2$, $HfO_2$, and $In_2O_3$ thin films with extrapolated Curie temperatures far in excess of 400 K [1, 2]. Magnetization of $TiO_2$ and $HfO_2$ films strongly decreases after 4 h annealing in oxygen and eventually disappears for 10 h annealing. Thus the authors of references [1, 2] concluded that oxygen **vacancies** are the main source of the magnetism in $TiO_2$ and $HfO_2$ thin films.

Striking phenomena such as the observation of room-temperature ferromagnetism in spherical nanoparticles (size 7–30 nm) of nonmagnetic oxides such as $CeO_2$, $Al_2O_3$, ZnO, $In_2O_3$, and $SnO_2$ have been reported [3]. These studies show that ferromagnetism is associated only with the nanoparticles, because the corresponding bulk samples are diamagnetic. Really, it was experimentally demonstrated that MgO nanocrystalline powders reveal room-temperature ferromagnetism, while MgO bulk exhibits diamagnetism [4]. The vacuum annealing of MgO nanocrystalline powders reduces ferromagnetism. The authors conclude that the ferromagnetism possibly originates from Mg vacancies at the surface and near the surfaces of nanograins. Large concentrations of Mg vacancies at the surfaces of nanograins can lead to magnetization via magnetic percolation [4].

The defects (impurities and vacancies) concentration increases near the sample surface, in particular allowing for the strong lowering of their formation energies [5, 6, 7]. Therefore, native vacancies should be present largely in the surface layer (more generally, in the vicinity of surface), and ferromagnetism observed in as grown nonmagnetic solids should arise primarily from the defects in the vicinity of surface.

In accordance with the numerous first-principles studies [5, 8, 9, 10, 11, 12, 13, 14] the origin of the magnetism and related properties in otherwise nonmagnetic materials is the cation defect only, i.e. induced by the magnetic ground state of the neutral cation vacancies that is represented by surrounding oxygen ions wave functions. It is obvious that some experimental results, namely the fact that oxygen annealing suppresses ferromagnetism in $HfO_2$, $SnO_2$ and $TiO_2$ films [1, 2, 3], demonstrating that anion oxygen vacancies play the main role in the appearance of ferromagnetism, seem to be in a disagreement with the aforementioned first principles calculations. From symmetry considerations we know that s-states cannot exist in the vicinity of surface. Quantum mechanical calculations performed earlier [15] confirmed the pure p-type one-electron impurity ground state at the surface, which was s-type in the bulk, so that some mixture of p- and s-states exists in the subsurface layers. The quantum-mechanical calculations [16] further show that the ground state of the impurities like He, $Li^+$, $Be^{2+}$ etc, as well as cation and anion vacancies in the binary solids is indeed the triplet (spin $\Sigma=1$) at the surface and close to the surface. Performed estimations have shown the possibility of appearance of ferromagnetic long-range order in the vicinity of surface.

Keeping in mind that magnetic properties depend essentially on the dielectric permittivity [15, 16], one has to expect many interesting peculiarities of the incipient ferroelectric properties due to the anomalous behavior of dielectric permittivity in a wide temperature range, what is typical for incipient ferroelectrics only. Note, that conventional ferroelectrics like $BaTiO_3$ and $PbTiO_3$ usually have a noticeable dielectric anisotropy below the Curie temperature, which is higher than room temperature. On the other hand the strong temperature and size dependence of incipient ferroelectric nanoparticles and thin films dielectric permittivity [17, 18] could lead to the interesting peculiarities in magnetic properties of the nanomaterials.

Ferromagnetic resonance (FMR) allows the measurement of the magnetic properties of a single particle in the powder and ceramic samples [3, 19, 20, 21, 22]. The FMR spectra contain information about the spontaneous magnetization and magnetic anisotropy of the sample and so the experiment can unambiguously reveal the presence of long-range magnetic order in the powder. Let us begin with consideration of the dielectric and magnetic properties of the single incipient ferroelectric nanoparticle.

## *2. Description of the model for the incipient ferroelectric nanoparticle properties calculations*

Using the quantum-mechanical approach combined with the image charge method we calculated the lowest energy levels of the impurities and neutral vacancies with two carriers (electrons or holes) localized the point $\mathbf{r}_0 = (0,0,z_0)$ near the surface z=0 of an incipient ferroelectric [see **Fig. 1a**]. We consider neutral vacancies with valency $Z = \pm 2$, which captured two carriers (electrons or holes), e.g. the $F^0$ center. For the case of the cation vacancy the cation atom should be added to form the perfect host lattice, and its two electrons should be localized at the nearest anions. As a result a negatively charged *defect* (-2e) with two holes appears in the continuous media. The situation is reversed for an anion vacancy: it can be modeled as a positively charged *defect* (+2e) with two electrons in its vicinity.

All calculations are based on the analytical results [16], obtained by the direct variational method applied to solve the Schrödinger-Vanjie equation in the effective mass approximation allowing for the defect-carriers Coulomb interaction along with the image charge contribution near the ideal surface. The approach, proposed much earlier [15], was used for the ground state calculations of the one-electron impurity center located near the flat surface.

Note, that the image charges method is developed for the continuum media approach, that requires the concept of the media dielectric permittivity $\varepsilon$ to describe the defect Coulomb potential. This is appropriate when the characteristic size of the carrier localization at the defect center is larger than the lattice constant [23, 24, 25, 26]. Since the defect is immovable the static dielectric permittivity should be used. Thus our calculations neglect the polaronic effects and dielectric anisotropy, i.e. hereinafter we used the effective static permittivity $\varepsilon \gg 1$ of the incipient ferroelectrics.

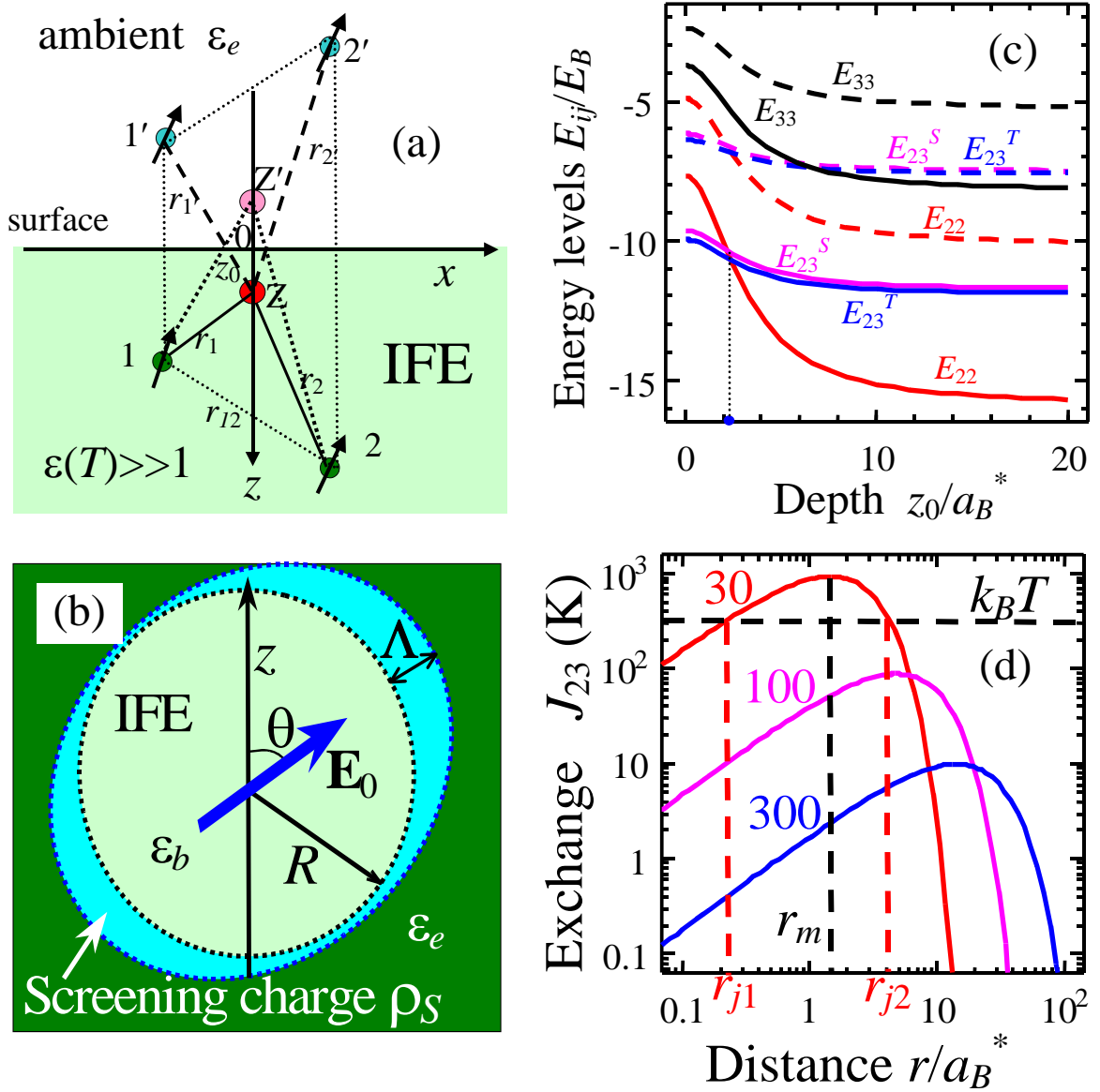

**Fig. 1.** (a) Defect at the distance $z_0$ under the surface of an incipient ferroelectric (IFE). Two carriers (electrons or holes) 1 and 2 (shown by green circles with arrows) are localized near the defect with effective charge $Ze$ (shown by red circle). Carrier image charges are shown as 1′ and 2′, defect image is Z′. Impurity atoms, cation or anion vacancies are considered in the model as defects placed in a perfect host lattice. (b) Incipient ferroelectric nanoparticles surrounded by the screening charge and matrix with dielectric permittivity $\varepsilon_e$. (c) The dependence of the energy levels $E_{ij}$ (in K) of the impurity or defect located at the distance $z_0$ from the surface calculated for effective mass $\mu=m_e$, $|Z|=2$, permittivity $\varepsilon=240$ (solid curves), 300 (dashed curves) corresponding to the dielectric permittivity of KTO and STO at room temperature. (d) Dependence of the exchange integral $J$ vs. the distance $r$ between the magnetic defects for permittivity $\varepsilon=30, 100, 300$ (figures near the curves).

The energy levels were calculated in the framework of the conventional perturbation theory as $E_{nl} = \langle \varphi_{nlm} | \hat{H} | \varphi_{nlm} \rangle$, where the coordinate dependence of the one-fermion (electron or hole) trial wave functions $\varphi_{nlm}$ was chosen in the form of hydrogen-like atom eigen functions ($n$ is the main quantum number, $l$ is the orbital quantum number, m is the magnetic quantum number, see **Appendix A.3**). The wave functions should be almost zero in the vacuum or air ambient (allowing for the high barrier determined by the work function at the solid/ambient interface), i.e. they satisfy the boundary condition $\varphi_{nlm}(x, y, z=0) = 0$. The lowest hydrogen-like atom wave functions, which satisfy the boundary conditions correspond to the $2p_z$ state: $\varphi_{210}(\mathbf{r}) = A(\alpha, z_0) \exp(-\alpha|\mathbf{r} - \mathbf{r}_0|) z$, where $\alpha$ is a variational parameter, and $A(\alpha, z_0)$ is the normalization constant. The next excited state is the $3p_z$ state: $\varphi_{310}(\mathbf{r}) = B(\beta, z_0) z (b(z_0) - \beta|\mathbf{r} - \mathbf{r}_0|) \exp(-\beta|\mathbf{r} - \mathbf{r}_0|)$, where $B(\beta, z_0)$ is the normalization constant, $\beta$ is the variational parameter. Here the variational parameter $b(0) = 2$ and $b(z_0 \to \infty) \to 1$. The variational parameters are determined from the energy minimum in the first order of the conventional perturbation theory, where the carrier-carrier Coulomb interaction and all interactions with the image charges are considered as perturbations.

The dependences of the energy levels $E_{ij}$ on the distance $z_0$ from the surface of incipient ferroelectrics KTaO$_3$ (KTO) and SrTiO$_3$ (STO) are shown in **Fig. 1c.** The magnetic triplet state $E_{23}^T$ appeared to be the ground state of the impurities and neutral vacancies in the vicinity of the incipient ferroelectric surface (**magnetic shell**), while the nonmagnetic singlet $E_{22}$ is the ground state below the distances $2 a_B^*$ (**nonmagnetic core**). Here $a_B^*(T, R) = (\varepsilon_e + \varepsilon(T, R)) 2\pi\varepsilon_0 \hbar^2 / (|Z|\mu e^2)$ is the effective Bohr radius, $\varepsilon(T, R)$ is the temperature and size dependent dielectric permittivity of incipient ferroelectric nanoparticles of radius $R$, $\mu$ is the fermion effective mass, $\varepsilon_e$ is the effective permittivity of the particle ambient, $\varepsilon_0$ is the universal dielectric constant.

When the energy level differences are small (see also **Fig. 1c**) the magnetic triplet and nonmagnetic singlet levels should be occupied with equal probabilities (25%) in a wide temperature range. We obtained that the following Pade approximations for the lowest energy levels z-dependences [16]:

$$E_{ij}^m(z_0) = \frac{E_{ij}^m(0) - E_{kl}^m(\infty)}{1 + 0.15 \cdot (\alpha(z_0) + \beta(z_0))^2 \, z_0^2 / a_B^{*3}} + E_{kl}^m(\infty). \tag{1}$$

Here subscripts $ij = 22, 23, 33$ correspond to the lowest levels at the surface and $kl = 1s1s, 1s2s, 2s2s$ correspond to the lowest levels far from the surface). The superscript $m = S, T$ indicates the singlet or triplet state. The localization radii are $\alpha(z_0) = \dfrac{\alpha(0) - 0.5 a_B^*}{1 + z_0^2/a_B^{*2}} + 0.5 a_B^*$ and $\beta(z_0) = \dfrac{\beta(0) - a_B^*}{1 + z_0^2/a_B^{*2}} + a_B^*$.

The special feature of nanosized incipient ferroelectrics is the strong temperature and size dependence of their dielectric permittivity [17, 18]. In particular, we derived the Barrett-type formula [27] for the temperature dependence of the dielectric permittivity of incipient ferroelectric nanospheres:

$$\varepsilon(T, R) = \varepsilon_b + \frac{1}{\varepsilon_0} \left( \alpha_T \left( \frac{T_q}{2} \coth\left(\frac{T_q}{2T}\right) - T_0 \right) + \frac{4\sigma_S(Q_{11} + 2Q_{12})}{R} + \frac{\Lambda \varepsilon_0^{-1}}{\varepsilon_e R + \Lambda(\varepsilon_b + 2\varepsilon_e)} + \frac{g}{(\lambda + \sqrt{3 g \varepsilon_0 \varepsilon_b}) R}\left(1 - \frac{3\Lambda \varepsilon_b}{\varepsilon_e R + \Lambda(\varepsilon_b + 2\varepsilon_e)}\right) \right)^{-1} \quad (2)$$

Here $R$ is the sphere radius, $T_q$ is the quantum oscillation temperature, $T_0$ is the virtual Curie temperature (the condition $T_q/2 > T_0$ must hold for incipient ferroelectric), $T$ is the absolute temperature (see **Fig. 1b**), $\Lambda$ is either the screening length ($\sim$ 1-10 nm) or the free-bound charges separation distance ($\sim$ 0.1 nm) at the particle surface (see **Appendix A.1-2** for details). Renormalization of the bulk permittivity originates from the extrinsic contribution of the surrounding charges such as the intrinsic surface stress (the term, proportional to $\sigma_S/R$), incomplete screening of the depolarization field outside the particle (the term, proportional to $\Lambda/(\varepsilon_e R + \Lambda(\varepsilon_b + 2\varepsilon_e))$), the intrinsic size effects related with polarization gradient g and the finite extrapolation length $\lambda$ (the term, proportional to $\dfrac{g}{R(\lambda + \sqrt{3 g \varepsilon_0 \varepsilon_b})}$).

Eq.(2) is valid in a wide temperature interval including low (quantum) temperatures [18]. Thus, in all the calculations below we use Eq.(2) with material parameters of KTO and STO defined in the **Table 1**.

**Table 1.** KTaO$_3$ and SrTiO$_3$ material parameters and LGD free energy expansion coefficients.

| Parameter | Unit | Materials | |
|---|---|---|---|
| | | **KTaO$_3$** [28, 29] | **SrTiO$_3$** [30] |
| Background permittivity $\varepsilon_b$ | dimensionless | 4 – 48 | 3 – 43 |
| LGD-expansion coefficient $\alpha_T$ | $10^6$ m/(F K) | 2.02 | 1.66 |
| Curie temperature $T_0$ | K | 13 | 36 |
| Quantum oscillation temperature $T_q$ | K | 55 | 100 |
| Electrostriction coefficient $Q_{11}$ | m$^4$/C$^2$ | 0.087 | 0.051 |
| Electrostriction coefficient $Q_{12}$ | m$^4$/C$^2$ | -0.023 | -0.016 |

| Intrinsic surface stress (surface tension) coefficient $\sigma_S$ | J/m$^2$ | 1 – 50 [31] | 1 – 50 |
|---|---|---|---|
| Screening length $\Lambda$ | nm | 0.2 – 4 | 0.2 – 4 |
| Ambient permittivity $\varepsilon_e$ | dimensionless | 1 – $\infty$ | 1 – $\infty$ |
| Extrapolation length $\lambda$ | nm | 0 – 1000 | 0 – 1000 |
| LGD-gradient coefficient $g$ | 10$^{-10}$ V·m$^3$/C | 1 – 5 [32] | 1 – 5 |
| LGD-expansion coefficient $\beta_{11}$ | 10$^9$ m$^5$/(C$^2$F) | 5 | 8.1 |
| LGD-expansion coefficient $\beta_{12}$ | 10$^9$ m$^5$/(C$^2$F) | 10 | 2.4 |
| LGD-expansion coefficient $\gamma_{111}$ | 10$^{12}$ m$^9$/(C$^4$F) | 4 | not found |
| Elastic compliances $s_{11}$ | 10$^{-12}$ m$^2$/N | 2.70 | 3.89 |
| Elastic compliances $s_{12}$ | 10$^{-12}$ m$^2$/N | -0.63 | -1.06 |
| Elastic compliances $s_{44}$ | 10$^{-12}$ m$^2$/N | 9.17 | 8.20 |
| Elastic stiffness $c_{11}$ | 10$^{11}$ N/m$^2$ | 4.31 | 3.36 |
| Elastic stiffness $c_{12}$ | 10$^{11}$ N/m$^2$ | 1.30 | 1.07 |
| Elastic stiffness $c_{44}$ | 10$^{11}$ N/m$^2$ | 1.09 | 1.27 |
| Effective mass of the localized carriers $\mu$ | free electron mass $m_e$ | 0.5 – 1 [33] | 0.5 – 1 [34] |

The temperature dependences of the KTO (a) and STO (b) dielectric permittivity $\varepsilon(T,R)$ are shown in **Figs. 2a,b**. The corresponding temperature dependences of the effective Bohr radius $a_B^*(T,R)$ are shown in **Figs. 2c,d.** It is seen from the figures that dielectric permittivity and the characteristic size of the carrier localization $r_d \cong a_B^*(T,R)$ decreases with decreasing particle radius due to the relation between $a_B^*(T,R)$ and $\varepsilon(T,R)$ (see Eq.(1)). The plateau that appears at low temperatures corresponds to the saturation of the Barrett term $\left(\frac{T_q}{2}\coth\left(\frac{T_q}{2T}\right) - T_0\right)$ at temperatures $T < T_0$.

Finite size effects make the incipient ferroelectric nanoparticles less sensitive in their dielectric susceptibility behavior (compare curves for different radii). However the permittivity $\varepsilon(T,R)$ in 2-20 nm sized nanoparticles is still much higher in comparison with the typical values for oxide materials and semiconductors. Due to the high values of $\varepsilon(T,R)$ the radius $a_B^*(T,R) > 5$ nm is much higher than the lattice constant $a = 0.4$ nm, providing the validity of the effective mass approximation as well as the self-consistent background for the introduction of dielectric permittivity in the continuous media approach [35].

The dependences of the KTO and STO energy levels of oxygen vacancies in the vicinity of surface on temperature and particle radius are shown in **Figs. 3,a,b** and **c,d** correspondingly. Magnetic triplet $E_{23}^T$ is the lowest one, but the energy level differences are very small. Thus the magnetic triplet and nonmagnetic singlet levels $E_{23}^S$, $E_{22}$ and $E_{33}$ could be occupied with approximately equal probabilities (25%) at temperatures $k_B T < E_{ij}(0)$. At high temperatures $k_B T \gg E_{ij}(0)$ the level occupation rapidly decreases, since the spontaneous thermal ionization starts. The dashed line $E_{ij}(0) = k_B T$ separates these regions. The difference between the energy levels slightly increases with increasing temperature (see plots **a,b** for fixed radius) and decreases with radius increase (see plots **c,d** for fixed temperature). In fact, the energy levels scale quasi-linearly with the inverse particle radius: $E_{ij}(0,R) \sim 1/R$ at small *R*.

Note, that both the energy levels depth and slope increase with the increasing of the screening length $\Lambda$ (compare solid and dotted curves). The explanation of this fact is that extrinsic size effects related with depolarization fields are significantly stronger at poor screening i.e. they scales as $\Lambda/(\Lambda + R)$. Moreover, the intrinsic size effects related with the polarization gradient are significantly stronger at small extrapolation length $\lambda$ than at high $\lambda$ (see Eq.(2)), i.e. they scales as $1/\lambda$ at high $\lambda$. The energies are proportional to the effective mass value: $E_{ij}(0,R) \sim \mu$. For simplicity we thus show only the case $\mu=m_e$ in the plots hereinafter. The low concentration of carriers in incipient ferroelectrics allows us to neglect their indirect interactions and correlation effects.

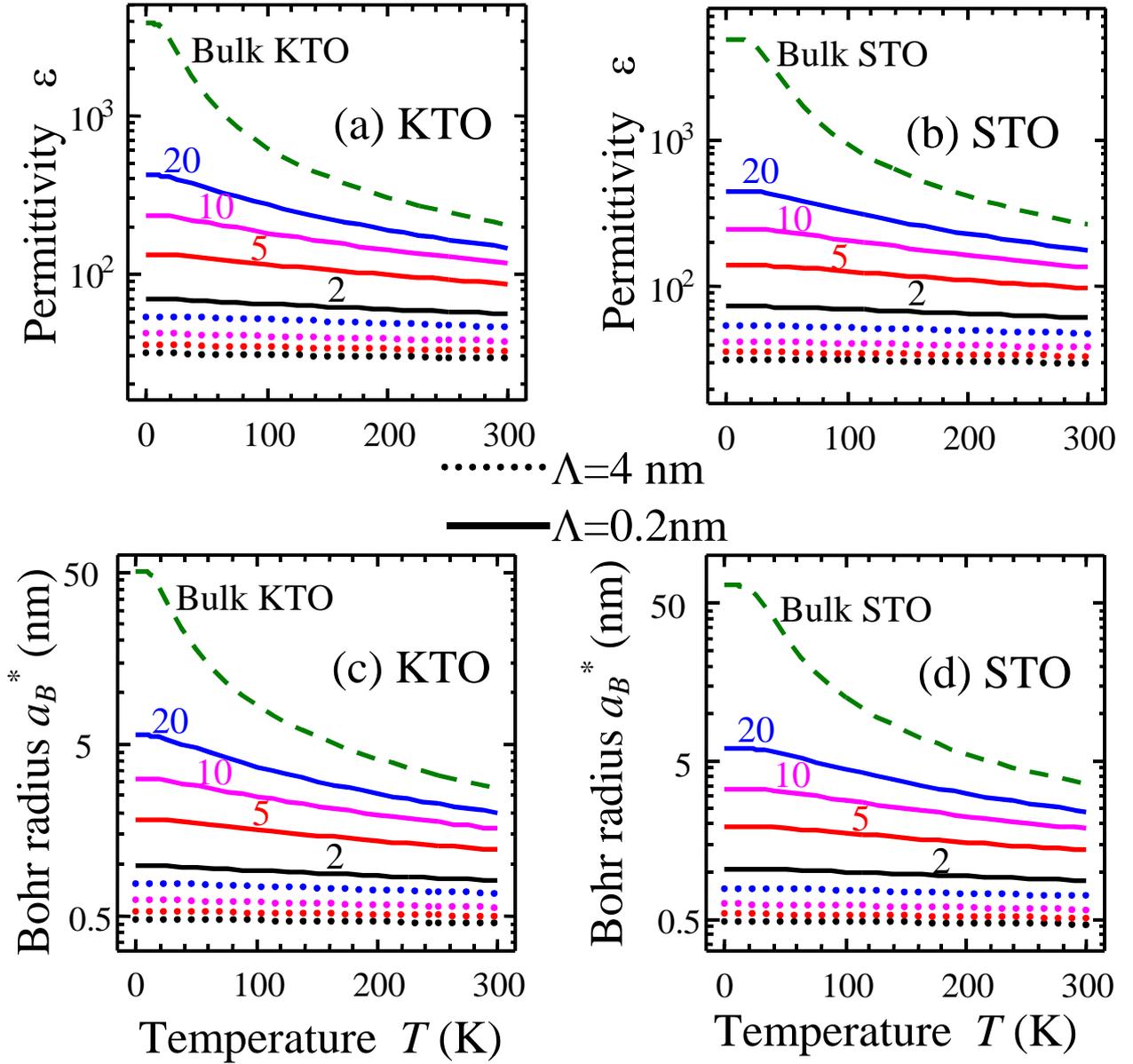

**Fig. 2**. Temperature dependences of the dielectric permittivity $\varepsilon(T, R)$ (a,b) and effective Bohr radius $a_B^*(T, R)$ (c,d) calculated for KTO and STO nanoparticles of different radii $R$ = 2, 5, 10, 20 nm (figures near the solid curves), screening lengths $\Lambda$ = 0.2 nm (solid curves from top to bottom), $\Lambda$ = 4 nm (dotted curves from top to bottom). Dashed curves correspond to the bulk material ($R \to \infty$). Extrapolation length $\lambda$ = 1 nm, effective mass $\mu = m_e$, gradient coefficient g = $10^{-10}$ V·m$^3$/C, surface tension coefficient $\sigma_S$ = 3 J/m$^2$, background permittivity $\varepsilon_b$=10, ambient permittivity $\varepsilon_e$=1−5.

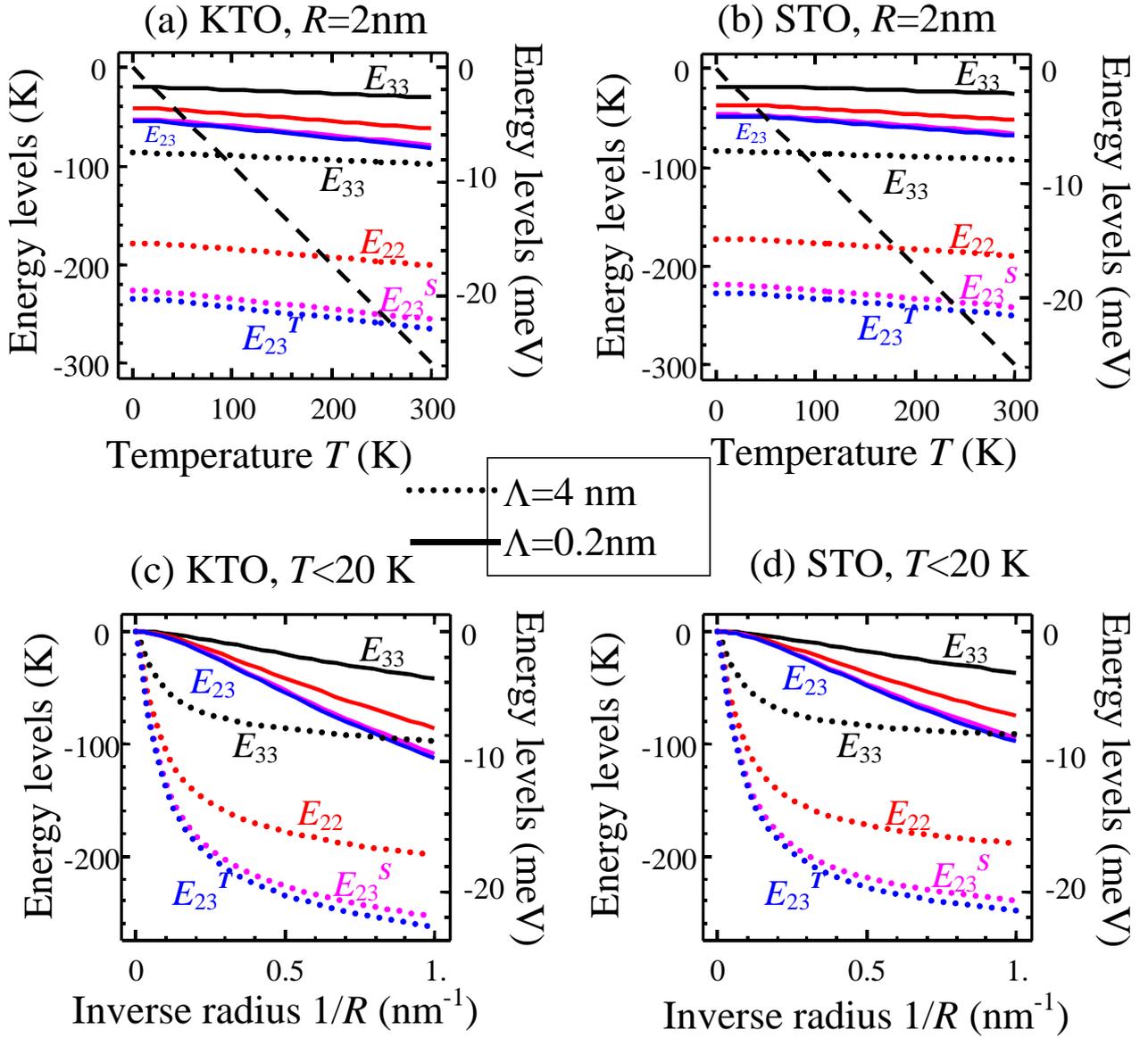

**Fig. 3**. Temperature (a,b) and radius (c,d) dependences of the lowest energy levels $E_{ij}(0)$ at the surface of KTO (a,c) and STO (b,d) calculated for screening lengths $\Lambda = 0.2$ nm (solid curves), $\Lambda = 4$ nm (dotted curves). Particle radius $R = 2$ nm for plots (a,b) and temperature $T < 20$ K for plots (c,d). Other parameters are the same as in Fig. 2.

Since the influence of the surface on the concentration of oxygen vacancies increase is noticeable, they act as a new type of randomly distributed magnetic defects appearing in the vicinity of surface.

## 3. Diagrams of magnetic phases induced by oxygen vacancies. Size-dependent ferromagnetism in the nanosized incipient ferroelectric nanoparticles

In what follows we will discuss the possible magnetic phase diagram induced by oxygen vacancies in the vicinity of the surface of incipient ferroelectrics STO and KTO.

Taking into account the random distribution of oxygen vacancies it is reasonable to operate with an average distance $r$ between them. Keeping in mind that the thickness of magnetic shell, where the existence of magnetic spins can be expected, is about $2a_B^*$, the relation between the concentration $N$ of the random magnetic defects and average distance between them $r$ could be calculated as $N = 6/\pi r^3$ if $r < 2a_B^*$ and $N = 4/(\pi r^2 a_B^*)$ if $r > 2a_B^*$ [26].

Note, that the defect-induced long-range ferromagnetic order can have a percolation nature [4], especially, when the problem dimensionality is reduced by the spatial confinement.

The dependence of the exchange integral $J_{23}(r)$ on the distance $r$ between the defects is shown in **Fig. 1d**. Note, that indexes "23" in the exchange integral denote that it was calculated on the triplet wave functions as $J_{23}(r) = \langle \varphi_{210}(\mathbf{r}_1)\varphi_{310}(\mathbf{r}_2)\tilde{V}_{12}(\mathbf{r}_1,\mathbf{r}_2)\varphi_{210}(\mathbf{r}_1)\varphi_{310}(\mathbf{r}_2) \rangle$. The exchange integral $J_{23}(r)$ is positive independently of the distance $r$. $J_{23}(r)$ tends to zero at $r \to 0$ (as calculated with $p$-states wave function) and has a pronounced maximum at distances $r = r_m$ and then vanishes exponentially with the distance increase $r \gg r_m$.

The exchange integral depends on the particle radius and temperature via the size and temperature dependence of dielectric susceptibility $\varepsilon(T,R)$. Namely, both our numerical and analytical calculations showed that $J_{23}(r) = \dfrac{J_m}{(\varepsilon(T,R)+\varepsilon_e)^2} f\left(\dfrac{r}{(\varepsilon_e + \varepsilon(T,R))a_B}\right)$, where the amplitude $J_m$ is virtually independent of the dielectric permittivity at $\varepsilon(T,R) \gg 1$ and $J_m/k_B \approx 858907$ K; $a_B = \dfrac{2\pi\varepsilon_0 \hbar^2}{|Z|\mu e^2}$ is the "vacuum" Bohr radius. The positive dimensionless function $f$ has the following approximate form $f(x) \approx \dfrac{x}{4}\exp\left(1-\dfrac{x}{4}\right)$. The maximal value $f(r_m) = 1$, where $r_m \approx 4(\varepsilon_e + \varepsilon(T,R))a_B = 4a_B^*(T,R)$.

Note, that two roots ($r_{j1} < r_m < r_{j2}$) of equation $J_{23}(r) = k_B T$ may exist at fixed temperature and particle radius $R < R_c(T)$. These roots tend to $r_m$ when the particle radius tends to the critical radius $R_c(T)$. An approximate expression for the critical radius $R_c(T)$ can be derived in the following way. Since $\varepsilon(T,R) \gg 1$ for incipient ferroelectrics, the condition $J(r_m) = k_B T$ determines the ferromagnetic long-range order phase boundary $R_c(T)$ of the diagram in the coordinates particle

radius – temperature. Using Eq.(2), under the condition $\varepsilon_e R \gg \Lambda(\varepsilon_b + 2\varepsilon_e)$ we have found the following approximation for the critical radius as:

$$R_c(T) \approx \frac{1}{\alpha_T T_0}\left(4\sigma_S(Q_{11}+2Q_{12}) + \frac{\Lambda}{\varepsilon_e \varepsilon_0} + \frac{g}{\lambda + \sqrt{3g\varepsilon_0 \varepsilon_b}}\right) \times \left(\frac{(\varepsilon_0 \alpha_T T_0)^{-1}}{\sqrt{J_m/k_B T} - \varepsilon_e - \varepsilon_b} - \left(\frac{T_q}{2T_0}\coth\left(\frac{T_q}{2T}\right) - 1\right)\right)^{-1} \quad (3)$$

The temperature dependence of the critical radius is shown in **Figs. 4**. It is seen that the radius strongly decreases with the temperature increase. Its value depends on the screening length and surface tension.

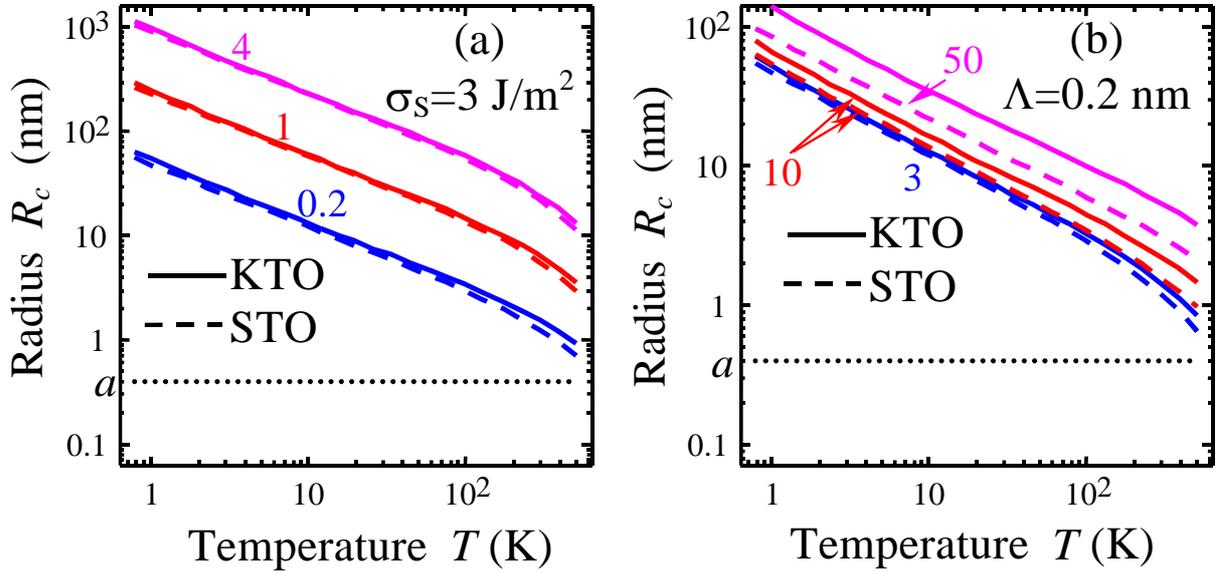

**Fig. 4**. Temperature dependence of the critical radius calculated for KTO (solid curves) and STO (dashed curves) material parameters, (a) surface tension coefficient $\sigma_S = 3$ J/m², screening length $\Lambda = 0.2$, 1, 4 nm (figures near the curves) and (b) $\Lambda = 0.2$ nm, $\sigma_S = 3$, 10, 50 J/m² (figures near the curves). Dotted horizontal line denote the lattice constant value, $a = 4$ Å. Other parameters are the same as in Fig. 2.

Further considerations for randomly distributed magnetic spins ($S=1$) can be performed in a conventional way. However the magnetic properties of incipient ferroelectric nanoparticles are strongly size-dependent due to the size dependence of their dielectric permittivity $\varepsilon(T, R)$ and the effective Bohr radius $a_B^*(T, R) \sim \varepsilon(T, R)$. The interaction energy between the magnetic defects "*i*" and "*j*" is:

$$J(r_{ij}, R, T) \approx \begin{cases} J_{23}(r_{ij}) = \dfrac{J_m}{(\varepsilon(T,R)+\varepsilon_e)^2}\left(\dfrac{r_{ij}}{r_{ex}}\right)\exp\left(1-\dfrac{r_{ij}}{r_{ex}}\right), & r_{ij} \leq 12 a_B^* \text{ (direct exchange)} \\ J_{dd}(r_{ij}) = -\dfrac{\mu_B^2}{4\pi\mu_0}\left(\dfrac{3(\mathbf{s}_i \mathbf{r})(\mathbf{r}\mathbf{s}_j)}{r_{ij}^5} - \dfrac{\mathbf{s}_i \mathbf{s}_j}{r_{ij}^3}\right), & r_{ij} \gg 12 a_B^* \text{ (magnetic dipole – dipole)} \end{cases}$$

(4)

Here $r_{ij}$ is the distance between the defects, $r_{ex} = 4 a_B^*(T,R)$ is the exchange radius. Note, that the magnetic dipole-dipole interaction $J_{dd}(r_{ij})$ is anisotropic and very weak in comparison with the direct exchange at the moderate distances, since it has an order $J_{dd} < \dfrac{\mu_B^2}{4\pi\mu_0 a_B^3}$ that is 4-5 orders smaller than the direct exchange $J_{23}(r_{ij})$ [24]. However the direct exchange interaction exceeds the dipole-dipole one only at distances $r_{ij} < r_{ex} \ln\left(\dfrac{J_m r_{ex}^3 4\pi\mu_0 e}{(\varepsilon(T,R)+\varepsilon_e)^2 \mu_B^2}\left(\ln\left(\dfrac{J_m r_{ex}^3 4\pi\mu_0 e}{(\varepsilon(T,R)+\varepsilon_e)^2 \mu_B^2}\right)\right)^4\right)$, which can be estimated as $r_{ij} < (15-25) r_{ex}$.

Using the above expression for the exchange energy, we calculated the phase diagrams and magnetic properties of KTO and STO nanoparticles *assuming a negligibly small halfwidth of their radius distribution function* $\vartheta(R)$, which is modeled by the Dirac delta function: $\vartheta(R) = \delta(R - \langle R \rangle)$. Thus we put $\langle R \rangle \equiv R$ in **Figs. 5-7**. Note, that modern CVD-based nanotechnology allows sintering of nanoparticles assemblies with almost equal sizes [36].

Using the mean field approximation, one can find the critical point of the transition between the ferromagnetic (FM) and paramagnetic (PM) phases. In considered case it is determined from the condition of the exchange energy $J_{23}(r)$ being equal to the thermal energy, $J_{23}(r) = k_B T$, where $T$ (in Kelvins) is the actual temperature. In accordance with **Fig. 1c** and Eq.(3) there exist either two ($r_{j1} < r_m < r_{j2}$), or one ($r_{j1} = r_m = r_{j2}$) or none ($R < R_c(T)$) roots of this equation with respect to the average distance $r$ between the defects. These roots $r_{j1,2}$ determine the two branches of the critical concentration of magnetic percolation $N_{1,2}^P = 6/(\pi r_{j1,2}^3)$ if $r_{j1,2} < 2 a_B^*$ and $N_{1,2}^P = 4/(\pi r_{j1,2}^2 a_B^*)$ if $r_{j1,2} > 2 a_B^*$. The branches $N_1^P(R,T)$ and $N_2^P(R,T)$ merge together at the critical radius $R_c(T)$, since $r_{j1} = r_{j2}$ at $R = R_c(T)$.

It is seen from **Figs.5** that ferromagnetic long-range order appears in the region of defect spin concentrations $N_2^p < N < N_1^p$ and particle radii $R < R_c(T)$ since $J(r,R) > k_B T$ in this region. At the defect spins concentration $N > N_1^p$ and $R < R_c(T)$ the ferro-glass (FG) phase appears. In the FG phase state $J(r,R) < k_B T$ for the defect spins located at average distance, while there are defect spins with rather high exchange, $J(r,R) > k_B T$. The boundary of the FM-PM phases is $N_2^p(R,T)$ and boundary of the FM-FG phases is $N_1^p(R,T)$. At vacancy concentrations $N < N_2^p$ and radii $R < R_c(T)$ the short-range ferromagnetic order and so the glass state may appear. The boundary between FG and PM phases $R = R_c(T)$ means that for most of the spins $J(r,R_c) < k_B T$. For the particle radii $R > R_c(T)$ only the PM phase is possible.

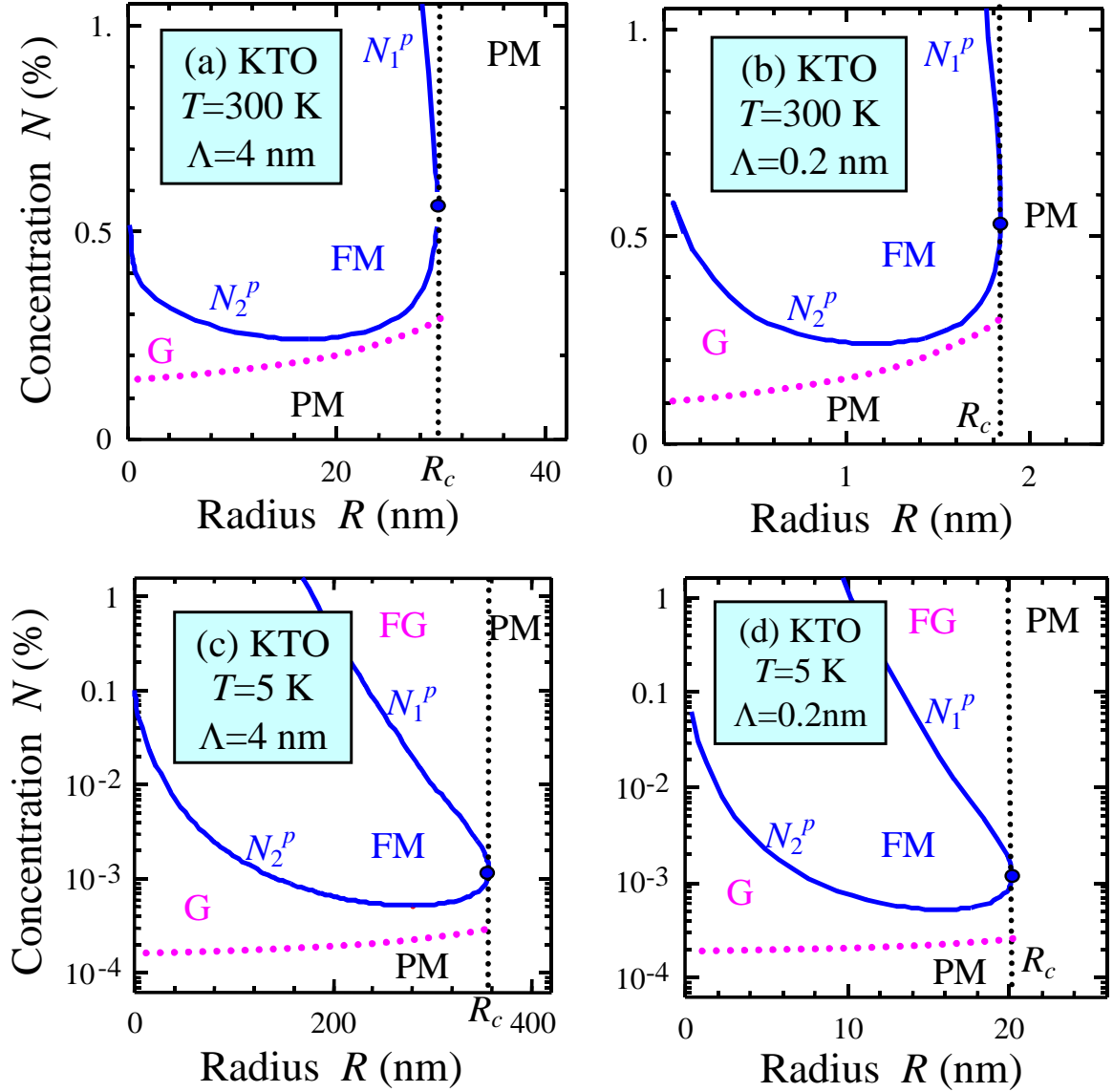

**Fig. 5**. Phase diagrams in coordinates "spin concentration-particle radius" calculated for material parameters of KTO, temperature $T=300$ K (a,b), 5 K (c,d); screening length $\Lambda = 0.2$ nm (b, d), $\Lambda = 4$ nm (a, c). G – glass state with diffuse boundary undetermined within our model, FG – ferroglass phase, FM – ferromagnetic phase, PM – paramagnetic phase. Other parameters are the same as in Fig. 2.

The surface-induced long-range ferromagnetism manifestation is not excluded at room temperatures, but here the nanoparticles radius should be less than 2-30 nm (see **Figs. 5a,b**). The concentration and radii region of FM and FG phases strongly increases with the temperature decrease (compare **Figs. 5c,d** at $T$=5 K with **Figs. 5a,b** at $T$=300 K). The concentration and radii region of FM and FG phases noticeably decreases with the screening length $\Lambda$ decrease (compare e.g. **Figs. 5a** at $\Lambda = 4$ nm with **Figs. 5b** at $\Lambda = 0.2$ nm).

Note, that the phase diagram for STO nanoparticles is similar to those of KTO ones. Possible phases are summarized in the **Table 2.**

**Table 2.** Size effect of the phase diagram of incipient ferroelectric nanoparticle

| Phase | Conditions | | |
|---|---|---|---|
| | **Particle radius $R$** | **Exchange energy $J$** | **Defect concentration $N$** |
| ferromagnetic phase (FM) | $R < R_c(T)$ | $J(r,R) > k_B T$ | $N_2^p < N < N_1^p$ |
| ferro glass phase (FG) | $R < R_c(T)$ | $J(r,R) < k_B T$ | $N > N_1^p$ |
| glass state (G) | $R < R_c(T)$ | $J(r,R) < k_B T$ | $N < N_2^p$ |
| paramagnetic phase (PM) | $R > R_c(T)$ | $J(r,R) < k_B T$ | $N > 0$ |

Up to now we actually considered a single nanoparticle. The super-paramagnetic state (SPM) can exist in the assembly of nanoparticles with small radii $R < r_{ex} < R_c(T)$ depending on temperature and other material parameters. The superparamagnetic phase could be fabricated on the basis of the composites with incipient ferroelectric nanoparticles in the ferromagnetic phase, induced by oxygen vacancies under the conditions described earlier. The surface of the particles has to be covered by some substance to prevent interaction between the particles with radius smaller than exchange interaction radius. The substance can contain screening charges as well (see **Fig. 1c**). The magnetic properties of such composite can be described similarly to the conventional superparamagnetic. Namely, the particle spin is equal to $\Sigma = \sum_i S_i$, where "$i$" numerates magnetic vacancies with spin $S_i$. Since the barrier for the particle spin $\Sigma$ reorientation is proportional to the particle volume $V$, the spin $\Sigma$ can rotate at $K_a V/k_B T < 1$ ($K_a$ is the anisotropy energy), so that its susceptibility behavior in the external magnetic field $H$ will be described with the Brillouin function, while in paramagnetic phase it will be the Langevin one. At temperatures lower than blocking temperature $T_B \sim U/k_B$ ($U$ is the energy barrier), magnetic hysteresis appears in the field dependence. The possibility to obtain SPM phase on the basis of ferromagnetic nanoparticles inserted into some other type host matrix without the magnetic ions is not excluded also.

Note, that composites, which contain particles in the mixed ferro-glass phase, can be suitable SPM phase observation, although the particle spin, $\Sigma$, will be smaller than for particles in the FM phase. The same is true for the volume *V*, that correspond to FM phase, so the condition of the free rotation can be softer, while the blocking temperature can be lower. Note, that in real nanocomposites the distribution of the particle sizes is not excluded as well as the coexistence of the mixture of nanoparticles in FM phase and FG state.

### 4. Magnetization and magnetic susceptibility

The developed approach allows analytical calculations of the magnetic properties of the considered system. In what follows we will demonstrate this on the examples of magnetization and susceptibility.

We performed analytical calculations of thedependences of the magnetization *M* and susceptibility $\chi = \frac{\partial M}{\partial H}$ on the magnetic field *H* for different particle sizes, temperature and defect concentration (see **Appendix B** for details). Below we described the results of our calculations.

The magnetization of a single nanosize particle is calculated as $M = \mu_B N \cdot s$, where *N* is the concentration of magnetic defects in the nanoparticle, $0 \le s \le 1$ is the fraction of coherently oriented spins, $\mu_B$ is the Bohr magneton.

We have found that the magnetization of incipient ferroelectric nanoparticles in the super-paramagnetic phase obeys the Brillouin law:

$$M(H) = N\mu_B \left( \frac{2\Sigma + 1}{2} \coth\left( \frac{2\Sigma + 1}{2} \frac{\mu_B H}{k_B T} \right) - \frac{1}{2} \coth\left( \frac{1}{2} \frac{\mu_B H}{k_B T} \right) \right) \qquad (5)$$

Here $\Sigma$ is the number of spins inside the nanoparticle, and $H(t)$ is the strength of the static or periodic magnetic field applied to the nanoparticle.

On the basis of the statistical physics approach [46], it is possible to derive the size and temperature dependence of the spontaneous magnetization in the FM phase, as:

$$M(H) = \mu_B N \tanh\left( \frac{1}{k_B T} \left( \mu_B H + J(T, R, r(N)) \frac{M}{\mu_B N} \right) \right). \qquad (6)$$

In Eq.(6) $J(T, R, r(N))$ is the exchange constant given by Eq.(3), after substituting the average distance $r(N)$ between magnetic defects. The temperature $T_c$ of the ferromagnetic phase transition is determined from the condition $J(T_c, R, r(N)) = k_B T_c$. At fixed temperature one can find either the critical radius or the defect concentration for this transition.

In the PM phase, where the defect spin (S=1) is able to rotate, we derived the magnetic field dependences of the magnetization induced by the *H*-field and the susceptibility as:

$$M = \mu_B N \left( \frac{\cosh\left(\frac{\mu_B H}{k_B T}\right) + \exp\left(\frac{-2J}{k_B T}\right)}{\sinh\left(\frac{\mu_B H}{k_B T}\right) + \exp\left(\frac{-2J}{k_B T}\right)\frac{\mu_B H}{k_B T}} - \frac{k_B T}{\mu_B H} \right) \quad (7a)$$

$$\chi = \mu_B N \left( \frac{k_B T}{\mu_B H^2} - \frac{1}{k_B T} \frac{1 + \exp\left(\frac{4J}{k_B T}\right) + \exp\left(\frac{2J}{k_B T}\right)\left(2\cosh\left(\frac{\mu_B H}{k_B T}\right) - \sinh\left(\frac{\mu_B H}{k_B T}\right)\frac{\mu_B H}{k_B T}\right)}{\left(\sinh\left(\frac{\mu_B H}{k_B T}\right)\exp\left(\frac{2J}{k_B T}\right) + \frac{\mu_B H}{k_B T}\right)^2} \right) \quad (7b)$$

Note, that Eq.(7b) is the modified Langevin-like law (see **Appendix B** for details).

The dependences of the magnetization $m = M/(\mu_B N)$ and susceptibility $\chi = \partial M/\partial H$ on the applied magnetic field $H$ as calculated for a single KTO nanoparticle in SPM (where $\chi(H)$ obeys the Brillouin law at temperatures more than $T_B$) and PM phases are shown in **Fig. 6** for low and room temperatures. Here we assume a negligibly small halfwidth of the particle radius distribution function. The magnetization saturation field increases with the temperature decrease, since the scaling factor $H/k_B T$ in Eqs.(5,7) decreases with temperature (compare **Fig. 6a,b**). The magnetization $M(H)$ of incipient ferroelectric nanoparticles of radius $R < r_{ex}(T)$ is described by the strongly nonlinear Brillouin-like law, while the susceptibility has a pronounced maximum at zero magnetic field as anticipated for the SPM phase (see solid curves in **Fig. 6**). At particle radii $R > R_c(T)$ the magnetization curves transform into the Langevin-like dependence, which is characteristic for the PM phase of non-interacting magnetic spins (see dashed curves in **Fig. 6**). With increase of the particle radius ($R >> R_c(T)$) the magnetization field dependence becomes linear, $M \sim H$, as expected for the paramagnetic state (see dashed line in **Fig. 6b**).

The estimated values of the coercive biases in the FM phase are unrealistically large (~100 Tesla) and above the "exchange field" $J/\mu_B$ or even anisotropy field [37]. Thus we calculated only the spontaneous magnetization $M(0)$ of the incipient ferroelectric nanoparticle. A pronounced spontaneous magnetization was obtained for the small particles with radius $r_{ex} < R < R_c(T)$ and magnetic spins concentrations $N_2^p < N < N_1^p$, which correspond to the FM phase (**Fig. 7**). At fixed particle radius the spontaneous magnetization decreases with the temperature increase and disappears at the critical temperature $T_c(R)$ (see **Fig. 7a,b**). The temperature $T_c(R)$ decreases with the increase of the particle radius (compare different curves in **Fig. 7a**) and $T_c(R)$ increases with the defect concentration (compare different curves in **Fig. 7b**). SPM phase occurs at $R \le r_{ex}(T)$ the. At fixed temperature the spontaneous magnetization decreases with increasing particle radius and disappears at the critical radius $R_c(T)$ of the FM to PM phase transition (see **Fig. 7c**). The radius $R_c(T)$ increases with the temperature decrease (compare different curves in **Fig. 7c**). **Fig. 7d** represents the magnetization as a function of ~ $k_B T/J$ scaled temperature.

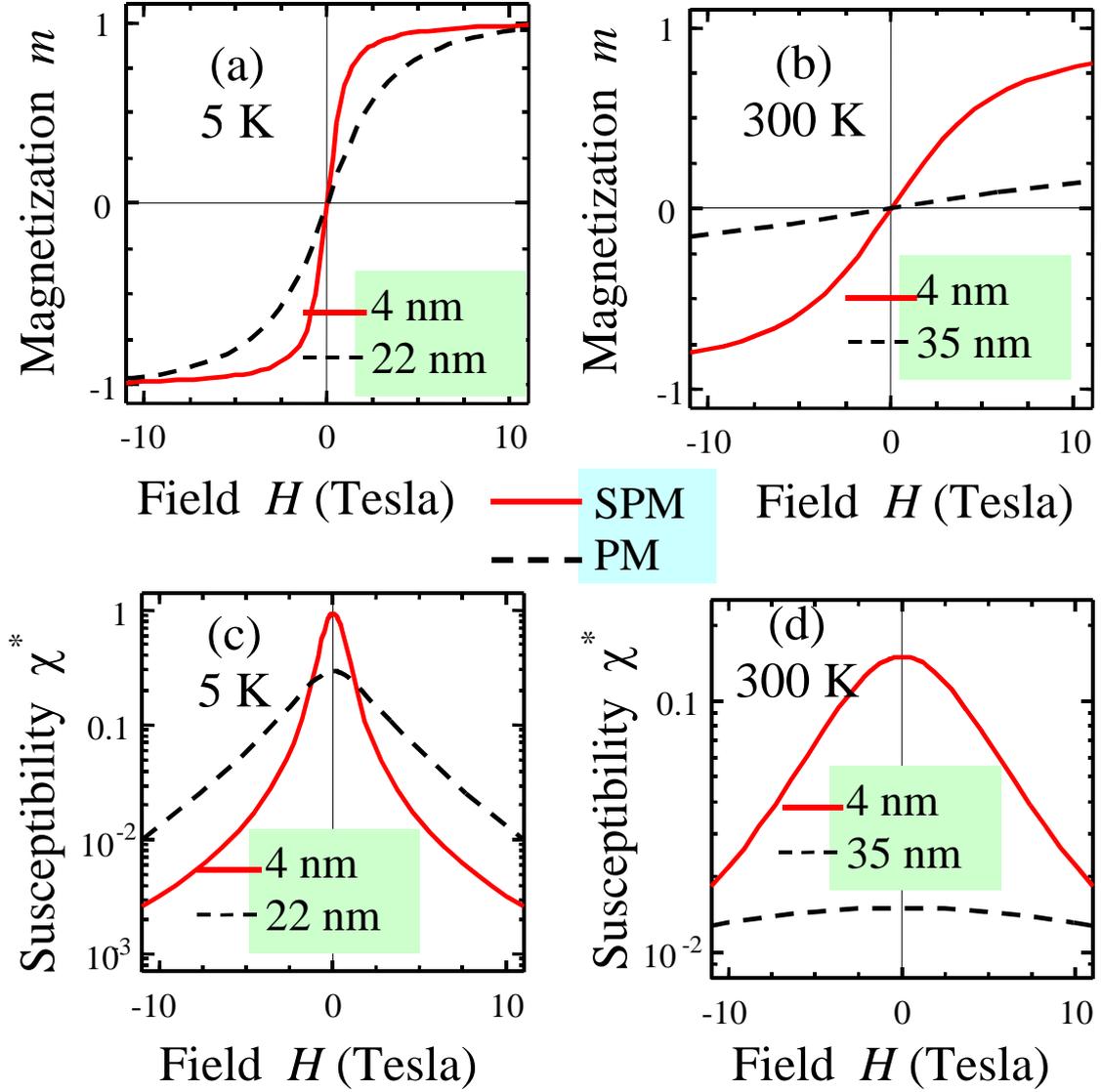

**Fig. 6.** Dependences of the magnetization $m = M/(\mu_B N)$ (a, b) and susceptibility $\chi^* = \chi \dfrac{k_B T}{\mu_B^2}$ (c, d) on the applied magnetic field $H$ calculated at temperatures T = 5 K and T = 300 K for KTO nanoparticles of radius $R$ = 4 nm (solid curve, SPM phase), 22, 35 nm (dashed curve, PM phase); screening length $\Lambda$ = 0.2 nm, for (a, c) and $\Lambda$ = 4 nm (b, d). Other parameters are the same as in Fig. 2.

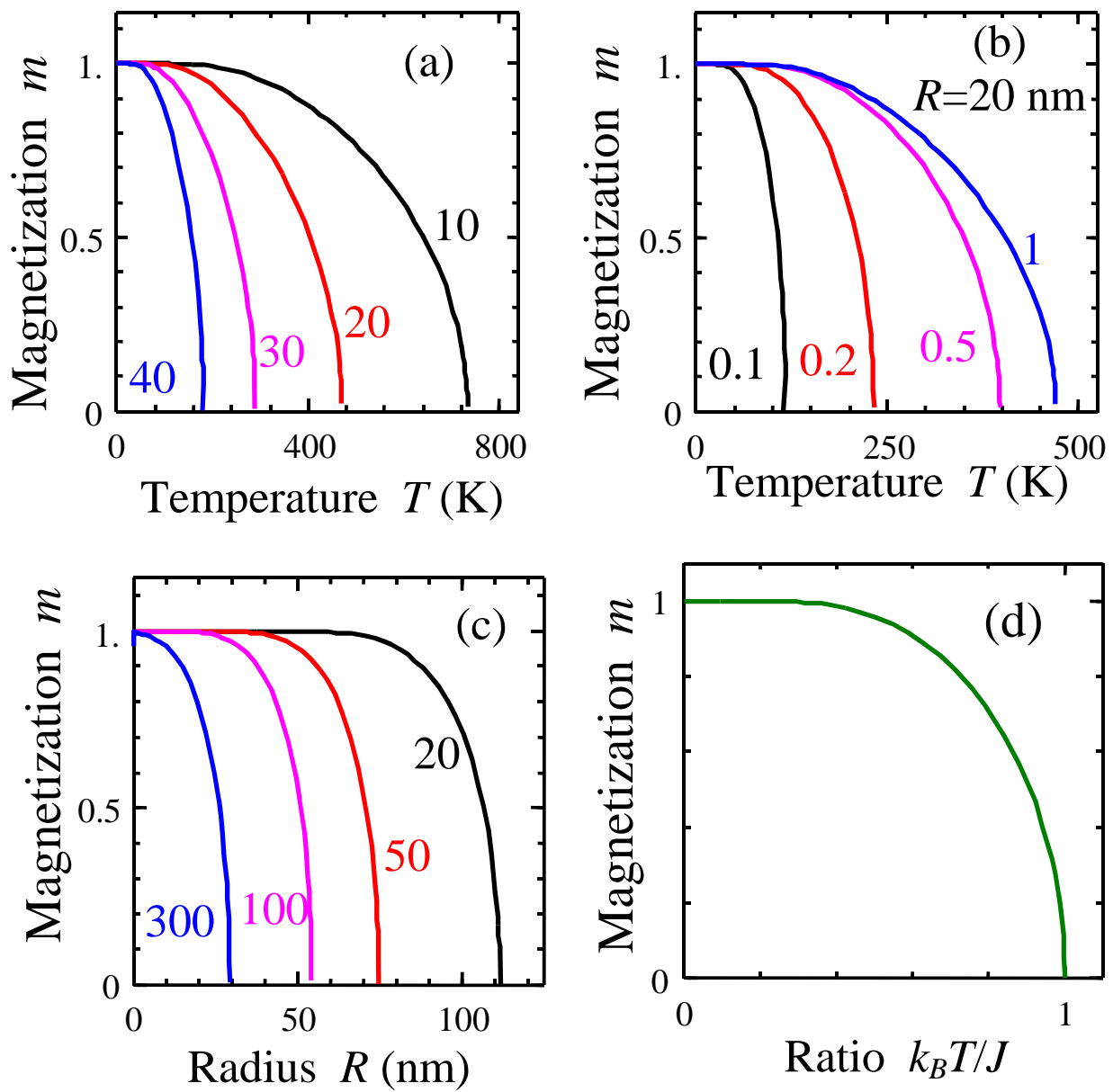

**Fig. 7.** Dependences of the normalized spontaneous magnetization $m = M/(\mu_B N)$ on temperature (a, b) and particle radius (c) calculated for KTO nanoparticles of radius $R$ = 10, 20, 30, 40 nm (figures near the curves) and fixed magnetic defects concentration $N$ = 1 % (a); magnetic defects concentration $N$ = 0.1, 0.2, 0.5, 1 % (figures near the curves) and radius $R$ = 20 nm (b); temperature $T$ =20, 50, 100, 300 K (figures near the curves) and $N$ = 1% (c). Plot (d) represents the magnetization as a function of scaled temperature ~$k_B T/J$. Screening length $\Lambda$ = 4 nm, other parameters are the same as in Fig. 2.

### 4. *Summary*

To summarize, we predict that undoped nanoparticles (size $\leq 10 - 100$ nm) of incipient ferroelectrics could become ferromagnetic up to the room temperatures due to the inherent presence of a new type of magnetic defects – oxygen vacancies, where magnetic triplet state is the ground state in the vicinity of surface (magnetic shell), while the nonmagnetic singlet is the ground state in the bulk (nonmagnetic core). The surface-induced magnetic states of the oxygen vacancies and other defects should exist at the surface of incipient ferroelectrics and below the surface up to distances of 5-50 nm.

The magnetic properties of nanoparticles were shown to be strongly size-dependent due to the size dependence of the incipient ferroelectric dielectric permittivity $\varepsilon(T,R)$ and the effective Bohr radius $a_B^*(T,R) \sim \varepsilon(T,R)$, which essentially influence exchange and magnetic dipole-dipole interactions. The ferromagnetic long-range order arises due to the percolation between the magnetic defects (vacancies). In particular, for particle radii $R$ less that the critical radius $R_c(T)$ the ferromagnetic long-range order appears in the shell region of thickness 5 – 50 nm once the concentration of magnetic defects $N$ exceeds the percolation concentration $N^p$. The critical radius $R_c(T)$ is calculated in the mean field theory from the condition $J(r, R_c, T) = k_B T$, where $J(r, R)$ is the exchange energy of the magnetic defects, and $r$ is the average distance between the defects. At vacancy concentrations $N < N^p$ (percolation is absent) and radii $R < R_c(T)$ a short-range ferromagnetic order and consequently a glass state may appear. For particle radii $R > R_c(T)$ only the paramagnetic phase is possible. The super-paramagnetic state appears in the assembly of non-interacting nanoparticles with narrow distribution function of their sizes, when the magnetic exchange radius becomes higher than the average particle size $\langle R \rangle$ and the halfwidth of the distribution function $\delta R \ll \langle R \rangle$. The magnetization of incipient ferroelectric nanoparticles in the super-paramagnetic phase is described by a Brillouin-like law.

We performed analytical calculations of the magnetization and susceptibility for different particle sizes, temperature and defect concentration. Pronounced spontaneous magnetization was obtained for small nanoparticles with radius $R < R_c(T)$ and magnetic spins concentrations $N > N^p$. With the nanoparticle radius increase $R > R_c(T)$ the magnetization curve transforms into a Langevin-like dependence, which is characteristic for the paramagnetic state of non-interacting magnetic spins.

**Appendix A**

**A.1. Extrinsic size effect contribution via the depolarization field due to the incomplete external screening. Polarization is homogeneous inside the nanoparticle**

Let us substitute the real shape of a given nanoparticle by an equivalent sphere of radius $R$. Firstly we calculate the depolarization field for the simplest case of a dielectrically isotropic core, shell and ambient materials. We consider a zero external field, since equations of electrostatics are linear and the corresponding solution for the sphere with shell in the homogeneous external field could be added to the solution found below [38].

The equations of state relating displacement **D**, electric field **E** and polarization **P** are:

$$\mathbf{D}_i \approx \mathbf{P} + \varepsilon_0 \varepsilon_b \mathbf{E}_i, \quad \mathbf{D}_s = \varepsilon_0 \varepsilon_s \mathbf{E}_s, \quad \mathbf{D}_e = \varepsilon_0 \varepsilon_e \mathbf{E}_e. \quad (A.1)$$

Here we used the so-called linearized model of ferroelectric nanoparticle core polarization and introduced its isotropic dielectric permittivity $\varepsilon_{11} = \varepsilon_{33} = \varepsilon_b$, where $\varepsilon_b$ is called background [39] or reference state permittivity [40]. External screening layer "s" has permittivity $\varepsilon_S$; ambient medium "e" has permittivity $\varepsilon_e$

Hereinafter we introduce the potential of electric field $\mathbf{E} = -\nabla \varphi(\mathbf{r})$. In spherical coordinates $\mathbf{r} = \{r, \theta, \varphi\}$ the potential inside each region $i$, $s$, $e$ acquires the form:

$$\varphi(r, \theta) = \begin{cases} \varphi_i(r, \theta), & 0 \leq r < R, \\ \varphi_s(r, \theta), & R \leq r < R_o, \\ \varphi_e(r, \theta), & r \geq R_o. \end{cases} \quad (A.2)$$

Nanoparticle radius is $R$, shell radius is $R_o$. Maxwell equation $\operatorname{div} \mathbf{D} = 0$ should be supplied with boundary conditions:

$$(\varphi_i - \varphi_s)\big|_{r=R} = 0, \quad (\mathbf{D}_s - \mathbf{D}_i)\mathbf{e}_r = \left(-\varepsilon_0 \varepsilon_s \frac{\partial \varphi_s}{\partial r} + \varepsilon_0 \varepsilon_b \frac{\partial \varphi_i}{\partial r} - P_3 \cos\theta\right)\bigg|_{r=R} = 0,$$

$$(\varphi_s - \varphi_e)\big|_{r=R_o} = 0, \quad (\mathbf{D}_e - \mathbf{D}_s)\mathbf{e}_r = \varepsilon_0 \left(-\varepsilon_e \frac{\partial \varphi_e}{\partial r} + \varepsilon_s \frac{\partial \varphi_s}{\partial r}\right)\bigg|_{r=R_o} = 0, \quad (A.3)$$

$$\varphi_i\big|_{r=0} < \infty, \quad \varphi_e\big|_{r \to \infty} = 0$$

Here $\mathbf{e}_r$ is the outer normal to the spherical surfaces.

As the first step we suppose that the polarization inside the sphere is homogeneous. The electrostatic potential inside the particle and screening layer satisfies Laplace equation $\Delta \varphi = 0$, while the media outside the particle may be semiconducting, its potential should satisfy the equation $\Delta\varphi_e - \varphi_e/l_d^2 = 0$ in Debye approximation with a screening length $l_d$.

The general solution of Laplace equation $\Delta\varphi = 0$, depending only on radius $r$ and polar angle $\theta$ is

$$\varphi(r,\theta) = \sum_{n=0}^{\infty}\left(a_n r^n + \frac{b_n}{r^{n+1}}\right) p_n(\cos\theta) \qquad (A.4)$$

where $a_n$ and $b_n$ are constants, $p_n$ are Legendre polynomials. One should leave in Eq. (A.4) only the terms with $n=1$ only. Thus we derived the solution as [41]:

$$\varphi_i(r,\theta) = a r \cos\theta, \quad 0 \le r < R. \qquad (A.5a)$$

Potential (18a) corresponds to a homogeneous field equal to -*a*.

$$\varphi_s(r,\theta) = \left(cr + \frac{b}{r^2}\right)\cos\theta, \quad R \le r < R_o, \qquad (A.5b)$$

$$\varphi_e(r,\theta) = \frac{d\exp(-r/l_d)(l_d + r) + f\exp(r/l_d)(r - l_d)}{r^2}\cos\theta, \quad r \ge R_o. \qquad (A.5c)$$

Potential (A.5c) corresponds to the field of a point dipole with moment ~*d*. Boundary conditions (A.3) give the system of linear equations for constants *a, b, c, d, f*. The electric field inside the incipient ferroelectric particle ($r < R$), is expressed via the effective depolarization factor $\eta$ as follows:

$$E_3 = -\frac{P_3}{\varepsilon_0 \varepsilon_b}\eta, \quad r < R. \qquad (A.6)$$

The effective depolarization factor $\eta$ essentially depends on the surroundings, namely:

(a) for the incipient ferroelectric particle in a semiconductor matrix the factor $\eta$ equals:

$$\eta = \frac{l_d(R+l_d)\varepsilon_b}{\left((R+l_d)^2 + l_d^2\right)\varepsilon_e + l_d(R+l_d)\varepsilon_b} \qquad (A.7a)$$

It is worth to note, that $\eta \approx \frac{\varepsilon_b}{\varepsilon_e}\frac{l_d}{R+l_d}$, for $R \gg l_d$.

(b) for the "incipient ferroelectric particle/ dielectric shell / dielectric matrix" the factor $\eta$ is

$$\eta = \frac{\left(R_o^3(\varepsilon_s + 2\varepsilon_e) - 2R^3(\varepsilon_e - \varepsilon_s)\right)\varepsilon_b}{R_o^3(\varepsilon_b + 2\varepsilon_s)(\varepsilon_s + 2\varepsilon_e) + 2R^3(\varepsilon_s - \varepsilon_b)(\varepsilon_e - \varepsilon_s)} \qquad (A.7b)$$

In the limiting case $\varepsilon_e \to \infty$ corresponding to the system "incipient ferroelectric particle/dielectric shell/conducting matrix" one has the following expression $\eta = \dfrac{(R_o^3 - R^3)\varepsilon_b}{R_o^3(\varepsilon_b + 2\varepsilon_s) + R^3(\varepsilon_s - \varepsilon_b)}$ and

$$\eta \approx \frac{\varepsilon_b}{\varepsilon_s}\frac{R_o - R}{R_o}, \quad \text{for} \quad R \gg R_o - R.$$

(c) the most general expression corresponds to the system consisting of incipient ferroelectric particle/ dielectric shell/semiconductor

$$\eta = \frac{\left(((R_o + l_d)^2 + l_d^2)(R_o^3 - R^3)\varepsilon_e + l_d(R_o + l_d)(R_o^3 + 2R^3)\varepsilon_s\right)\varepsilon_b}{\left((R_o + l_d)^2 + R_d^2\right)\left(R_o^3(\varepsilon_b + 2\varepsilon_s) + R^3(\varepsilon_s - \varepsilon_b)\right)\varepsilon_e + l_d(R_o + l_d)\left(R_o^3(\varepsilon_b + 2\varepsilon_s) - 2R^3(\varepsilon_s - \varepsilon_b)\right)\varepsilon_s} \quad (A.7c)$$

**A.2. Intrinsic size effect contribution via the depolarization field due to the incomplete external screening and inhomogeneous polarization inside the nanoparticle**

As the second step let us consider the case of inhomogeneous polarization inside the particle, supposing only a radial dependence, $P_z(r)$. Below we show that it is not rigorous, but in some cases this could be the first approximation.

*A.2a. The ideally screening ambient media.* Firstly we consider the particle inside the ideally screening ambient media ($l_d \to 0$) without a dielectric shell. It is obvious, that the solution of more complicated problems could be constructed by an appropriate combination of simpler solutions.

The electrostatic potential inside the particle satisfies Poisson equation

$$\Delta\varphi = \frac{div(\mathbf{P})}{\varepsilon_0 \varepsilon_b} \tag{A.8a}$$

For the case of $\mathbf{P} = (0, 0, P_3(r))$ (A.8a) reduces to

$$\Delta\varphi = \frac{\cos\theta}{\varepsilon_0 \varepsilon_b}\frac{\partial P_3(r)}{\partial r} \tag{A.8b}$$

Boundary conditions are

$$\varphi\big|_{r=0} < \infty, \quad \varphi\big|_{r=R} = 0 \tag{A.9}$$

It is natural to look for the solution of (A.9) in the form similar to (A.4), $\varphi(r,\theta) = \sum_{n=0}^{\infty} f_n(r) p_n(\cos\theta)$. Using the orthogonality of Legendre polynomials, one could see that only the term with *n*=1 will be sufficient. Thus, introducing the Ansatz $\varphi(r,\theta) = \psi(r)\cos\theta$ we obtain:

$$\frac{\partial^2 \psi(r)}{\partial r^2} + \frac{2}{r}\frac{\partial \psi(r)}{\partial r} - \frac{2}{r^2}\psi(r) = \frac{1}{\varepsilon_0 \varepsilon_b}\frac{\partial P_3(r)}{\partial r} \tag{A.10}$$

The solution of (A.10) could be found in the form

$$\psi(r) = \frac{1}{\varepsilon_0 \varepsilon_b} \left( \frac{1}{r^2} \int_0^r P_3(\tilde{r}) \tilde{r}^2 d\tilde{r} - \frac{r}{R^3} \int_0^R P_3(\tilde{r}) \tilde{r}^2 d\tilde{r} \right) \tag{A.11}$$

Now we can find the electric field $\mathbf{E} = -\nabla(\psi(r)\cos\theta)$. The z-component is:

$$E_3 = -\cos\theta \frac{\partial(\psi(r)\cos\theta)}{\partial r} + \frac{\sin\theta}{r} \frac{\partial(\psi(r)\cos\theta)}{\partial \theta} = -\cos^2\theta \frac{\partial(\psi(r))}{\partial r} - \sin^2\theta \frac{\psi(r)}{r} \tag{A.12}$$

It is seen that the z-component could be independent of $\theta$ in the very specific case $\psi(r) \sim r$ only (which also means that $E_3 = const$). That is why the supposition $P_3(r)$ is not rigorous. However, the approximate evident expression for $E_3$ follows from (A.11)-(A.12):

$$\begin{aligned} E_3 &= \frac{1}{\varepsilon_0 \varepsilon_b} \left( \frac{1}{R^3} \int_0^R P_3(\tilde{r}) \tilde{r}^2 d\tilde{r} - \cos^2\theta\, P_3(r) + \frac{2\cos^2\theta - \sin^2\theta}{r^3} \int_0^r P_3(\tilde{r}) \tilde{r}^2 d\tilde{r} \right) = \\ &= \frac{1}{\varepsilon_0 \varepsilon_b} \left( \frac{1}{R^3} \int_0^R P_3(\tilde{r}) \tilde{r}^2 d\tilde{r} - \frac{1}{3} P_3(r) - \frac{3\cos^2\theta - 1}{3r^3} \int_0^r \frac{\partial P_3(\tilde{r})}{\partial \tilde{r}} \tilde{r}^3 d\tilde{r} \right) \end{aligned} \tag{A.13}$$

It is seen that the first two terms are reduced to the form, proposed by us earlier[42] on the basis of the variation method, $E_3 \approx (\langle P_3(r) \rangle - P_3(r))/(3\varepsilon_0 \varepsilon_b)$. They are independent of $\theta$. The last term in (A.13) is proportional to the Legendre polynomial $p_2(\cos\theta)$ and corresponds to a divergent field like that of a dipole source. However, it has the impact only on the regions of the particle outside the range where the polarization changes rapidly. For example, if one has a particle with almost constant polarization throughout the particle except near a thin surface layer, $\partial P_3(r)/\partial r \approx 0$ at $0 < r < R - \delta R$, and surface layer with gradient polarization, $r\, \partial P_3(r)/\partial r \sim P_3(r)$ at $R - \delta R < r < R$, then the divergent term in (A.13) could be of order of the first two terms only in the surface layer $R - \delta R < r < R$.

*A.2b. The semiconducting ambient media.* For the case of a ferroelectric nanoparticle inside a semiconducting ambient media with a screening length $l_d$ let us look for the solution in the form:

$$\psi_i(r) = \frac{1}{\varepsilon_0 \varepsilon_b} \left( \frac{1}{r^2} \int_0^r P_3(\tilde{r}) \tilde{r}^2 d\tilde{r} - \frac{r}{R^3} \int_0^R P_3(\tilde{r}) \tilde{r}^2 d\tilde{r} \right) - E_i r, \quad r < R, \tag{A.14a}$$

$$\psi_e(r) = E_e \frac{\exp(-r/l_d)(l_d + r)}{r^2}, \quad r \geq R. \qquad \text{(see (A.5c))} \tag{A.14b}$$

Using the conditions

$$(\psi_i - \psi_e)|_{r=R} = 0, \quad \left( -\varepsilon_0 \varepsilon_e \frac{\partial \psi_e}{\partial r} + \varepsilon_0 \varepsilon_b \frac{\partial \psi_i}{\partial r} - P_3 \right)\bigg|_{r=R} = 0, \tag{A.15}$$

$$\psi_i|_{r=0} < \infty, \quad \psi_e|_{r \to \infty} = 0$$

it is possible to find the constants $E_i$ and $E_e$ and then to write the solution for the potential inside the nanoparticle in the form:

$$\psi_i(r) = \frac{1}{\varepsilon_0 \varepsilon_b} \left( \frac{1}{r^2} \int_0^r P_z(\tilde{r}) \tilde{r}^2 d\tilde{r} - \frac{r}{R^3} \int_0^R P_3(\tilde{r}) \tilde{r}^2 d\tilde{r} \right) +$$
$$+ r \frac{3}{R^3} \int_0^R P_3(\tilde{r}) \tilde{r}^2 d\tilde{r} \frac{1}{\varepsilon_0} \frac{l_d(R + l_d)}{\left((R + l_d)^2 + l_d^2\right) \varepsilon_e + l_d(R + l_d)\varepsilon_b}, \quad r < R \tag{A.16}$$

The electric field z-component inside the nanoparticle is

$$E_3 = \frac{1}{\varepsilon_0 \varepsilon_b} \left( \frac{1}{R^3} \int_0^R P_3(\tilde{r}) \tilde{r}^2 d\tilde{r} - \frac{1}{3} P_3(r) - \frac{3\cos^2\theta - 1}{3r^3} \int_0^r \frac{\partial P_3(\tilde{r})}{\partial \tilde{r}} \tilde{r}^3 d\tilde{r} \right) -$$
$$- \frac{1}{\varepsilon_0} \frac{3}{R^3} \int_0^R P_3(\tilde{r}) \tilde{r}^2 d\tilde{r} \frac{l_d(R + l_d)}{\left((R + l_d)^2 + l_d^2\right) \varepsilon_e + l_d(R + l_d)\varepsilon_b} \tag{A.17}$$

The last term in Eq.(A.17) is related to the non-ideal screening either due to the dead layer or finite screening length.

Now we could proceed with the consideration of the impact of the depolarization field on the particle dielectric properties. The LGD free energy density expansion for incipient ferroelectrics is

$$F = \begin{pmatrix} \frac{\alpha}{2} P^2 + \frac{\beta_{11}}{4} \left(P_1^4 + P_2^4 + P_3^4\right) + \frac{\beta_{12}}{2} \left(P_1^2 P_2^2 + P_2^2 P_3^2 + P_2^2 P_3^2\right) + \frac{\gamma_{111}}{6} \left(P_1^6 + P_2^6 + P_3^6\right) - \\ + Q_{ijkl} \sigma_{ij} P_k P_l - \frac{s_{ijkl}}{2} \sigma_{ij} \sigma_{kl} + \frac{g}{2} (\nabla \mathbf{P})^2 - \frac{1}{2} \mathbf{P} \mathbf{E}_d \end{pmatrix} \tag{A.18}$$

Let us consider one component of the polarization and suppose that effects of mechanical constraints are already included into the renormalized free energy coefficients. The equation of state is

$$\alpha P_3 + \beta P_3^3 - g\left(\frac{\partial^2 P_3}{\partial r^2} + \frac{2}{r} \frac{\partial P_3}{\partial r}\right) = E_3, \tag{A.19a}$$

with boundary conditions

$$\left. \left(\alpha_S P_3 + g \frac{dP_3}{dr}\right) \right|_{r=R} = 0, \tag{A.19b}$$

Neglecting the diverging depolarization field (i.e. the last term in Eq.(A.17)) for single-domain nanoparticles, the depolarization field (A.17) can be written in the form

$$E_3(r) \approx \frac{1}{3\varepsilon_0 \varepsilon_b} \left(\overline{P} - P_3(r)\right) - \frac{\eta}{\varepsilon_0 \varepsilon_b} \overline{P} \tag{A.20}$$

Here we introduced an average polarization of the particle $\overline{P} = \frac{3}{R^3} \int_0^R P_3(\tilde{r}) \tilde{r}^2 d\tilde{r}$. The depolarization factor η is defined in accordance with Eqs.(A.7).

As it was proposed earlier [43], let us look for the solution of Eq. (A.19) in the form $P_3(r) = \overline{P} + p(r)$ with the deviation $p$ regarded as small, $|p(r)| \ll |\overline{P}|$, and $\overline{p(r)} \equiv 0$. So, the linearized problem (A.19)-(A.20) acquires the form:

$$\begin{cases} \left(\alpha + 3\beta\overline{P}^2 + \dfrac{1}{3\varepsilon_0\varepsilon_b}\right)p - g\left(\dfrac{\partial^2 p}{\partial r^2} + \dfrac{2}{r}\dfrac{\partial p}{\partial r}\right) = -\left(\alpha\overline{P} + \beta\overline{P}^3\right) - \dfrac{\overline{P}}{\varepsilon_0\varepsilon_b}\eta, \\ \left.\left(\alpha_S p + g\dfrac{dp}{dr}\right)\right|_{r=R} = -\alpha_S \overline{P} \end{cases} \qquad (A.21)$$

The solution of the linear problem (A.4) has the form:

$$P_3(r) = \dfrac{2\beta\overline{P}^3 + \dfrac{\overline{P}}{\varepsilon_0\varepsilon_b}\left(\dfrac{1}{3} - \eta\right)}{\alpha + 3\beta\overline{P}^2 + \dfrac{1}{3\varepsilon_0\varepsilon_b}} s\!\left(r, R, \xi(\overline{P})\right), \qquad (A.22a)$$

where the space distribution function s is governed by:

$$s(r, R, \xi) = 1 - \dfrac{R^2}{R\lambda\cosh(R/\xi) + \xi(R-\lambda)\sinh(R/\xi)} \dfrac{\sinh(r/\xi)}{r/\xi}, \qquad (A.22b)$$

$$\xi = \sqrt{\dfrac{3\varepsilon_0\varepsilon_b g}{3\varepsilon_0\varepsilon_b\left(\alpha + 3\beta\overline{P}^2\right) + 1}}. \qquad (A.22c)$$

Here $\lambda = g/\alpha_S$ is an extrapolation length. The average polarization $\overline{P}$ should be determined self-consistently from the spatial averaging of Eq.(A.22), that leads to the following equation

$$\overline{P} = \dfrac{3\varepsilon_0\varepsilon_b\left(2\beta\overline{P}^3\right) + \overline{P}(1 - 3\eta)}{3\varepsilon_0\varepsilon_b\left(\alpha + 3\beta\overline{P}^2\right) + 1}\,\overline{s}(R,\xi), \qquad (A.23a)$$

$$\overline{s}(R,\xi) = \left(1 - \dfrac{3\xi^2(R - \xi\tanh(R/\xi))}{R(R\lambda + (R-\lambda)\xi\tanh(R/\xi))}\right) \xrightarrow{R \gg \xi} \left(1 - \dfrac{3\xi^2}{R(\lambda + \xi)}\right). \qquad (A.23b)$$

Allowing for the dependence of the characteristic length $\xi$ on the average polarization, Eq.(A.23) is a transcendental equation for $\overline{P}$. For the case of no external field, $E_0=0$, the equation (A.23a) could be rewritten in a more convenient way:

$$\overline{P} = \sqrt{\dfrac{-\alpha - \dfrac{1}{\varepsilon_0\varepsilon_b}\left(\dfrac{1}{3} - \eta\right)(1 - \overline{s}(R,\xi)) - \dfrac{\eta}{\varepsilon_0\varepsilon_b}}{\beta(1 + 2(1 - \overline{s}(R,\xi)))}}. \qquad (A.23c)$$

Since $\varepsilon_0\varepsilon_b\left(\alpha + 3\beta\overline{P}^2\right) \ll 1$ for most of ferroelectrics, the following approximation is valid with high accuracy:

$$\xi = \sqrt{\dfrac{3\varepsilon_0\varepsilon_b g}{3\varepsilon_0\varepsilon_b\left(\alpha + 3\beta\overline{P}^2\right) + 1}} \approx \xi_0 = \sqrt{3\varepsilon_0\varepsilon_b g} = const \qquad (A.24)$$

Using the approximation (A.24) the solution of equation (A.23c) could be written as:

$$\overline{P} = \sqrt{\frac{1}{\beta}\left(1+\frac{6\varepsilon_0\varepsilon_b g}{(\lambda+\xi_0)R}\right)^{-1}\left(-\alpha-\frac{3g}{R(\lambda+\xi_0)}\left(\frac{1}{3}-\eta\right)-\frac{\eta}{\varepsilon_0\varepsilon_b}\right)}$$

$$\approx \sqrt{\left(-\alpha-\frac{3g}{R(\lambda+\xi_0)}\left(\frac{1}{3}-\frac{\varepsilon_b\Lambda}{\varepsilon_e R+\Lambda(\varepsilon_b+2\varepsilon_e)}\right)-\frac{1}{\varepsilon_0}\frac{\Lambda}{\varepsilon_e R+\Lambda(\varepsilon_b+2\varepsilon_e)}\right)\frac{1}{\beta}\left(1+\frac{6\varepsilon_0\varepsilon_b g}{(\lambda+\xi_0)R}\right)^{-1}}.\quad (A.25)$$

Here $\Lambda$ is either the external screening length $l_d$ or the free-bound charges separation distance $R_o - R$ in agreement with Eqs.(A.7) In Eqs.(A.23b) and (A.25) we considered the limit $R>>\xi_0$, which is valid for most of the cases, since $\xi_0$ is usually of the order of a lattice constant.

Finally, allowing for the temperature and stress dependence of $\alpha$ for the incipient ferroelectrics, we derived from (A.23c) and (A.25) the renormalization of the expansion coefficient $\alpha$ as:

$$\alpha(T,R) = \alpha_T\left(\frac{T_q}{2}\coth\left(\frac{T_q}{2T}\right)-T_0\right) + \frac{4\sigma_S(Q_{11}+2Q_{12})}{R} + \frac{g(1-3\eta(R))}{(\lambda+\sqrt{3g\varepsilon_0\varepsilon_b})R} + \frac{\eta(R)}{\varepsilon_0\varepsilon_b}$$

$$\approx \alpha_T\left(\frac{T_q}{2}\coth\left(\frac{T_q}{2T}\right)-T_0\right) + \frac{4\sigma_S(Q_{11}+2Q_{12})}{R} + \quad\quad (A.26)$$

$$+\frac{g}{(\lambda+\sqrt{3g\varepsilon_0\varepsilon_b})R}\left(1-\frac{3\varepsilon_b\Lambda}{\varepsilon_e R+\Lambda(\varepsilon_b+2\varepsilon_e)}\right) + \frac{1}{\varepsilon_0}\frac{\Lambda}{\varepsilon_e R+\Lambda(\varepsilon_b+2\varepsilon_e)}$$

The last term in (A.26) is responsible for the extrinsic contribution to the size effect due to the imperfect screening[44] (also compare the last term in Eq.(A.26) with Eq.(3) from Ref.[45]).

### A.3. Wave functions and boundary condition

The boundary condition $\varphi_{nlm}(x,y,z=0)=0$ corresponds to an infinitely high barrier at the boundary solid-ambient medium (e.g. vacuum, atmosphere or semiconducting/electrolyte soft matter). The condition becomes nearly exact for the case of narrow-gap semiconductors and approximate for insulators. Two-fermion coordinate wave functions should be constructed from the functions $\varphi_{210}(\mathbf{r})$ and $\varphi_{310}(\mathbf{r})$ in the following way: singlet $\psi_{22}(\mathbf{r}_1,\mathbf{r}_2)=\varphi_{210}(\mathbf{r}_1)\varphi_{210}(\mathbf{r}_2)$, $\psi^S_{23}(\mathbf{r}_1,\mathbf{r}_2)=\frac{1}{\sqrt{2}}(\varphi_{210}(\mathbf{r}_1)\varphi_{310}(\mathbf{r}_2)+\varphi_{210}(\mathbf{r}_2)\varphi_{310}(\mathbf{r}_1))$ and $\psi_{33}(\mathbf{r}_1,\mathbf{r}_2)=\varphi_{310}(\mathbf{r}_1)\varphi_{310}(\mathbf{r}_2)$ with the full spin $\Sigma=0$ and triplet $\psi^T_{23}(\mathbf{r}_1,\mathbf{r}_2)=\frac{1}{\sqrt{2}}(\varphi_{210}(\mathbf{r}_1)\varphi_{310}(\mathbf{r}_2)-\varphi_{210}(\mathbf{r}_2)\varphi_{310}(\mathbf{r}_1))$ with full spin $\Sigma=1$. The energy difference between the lowest triplet and singlet states strongly depends on the carrier effective mass $\mu$, dielectric permittivity of the film $\varepsilon$ and the distance from the surface $z_0$. The triplet state $\psi^T_{23}$ should become the magnetic one ($s_z=\pm 1$) allowing for the Hund's rule that orient two fermions (electrons or holes) spins in the same direction.

**Appendix B**

The partition function of the system could be written as (see e.g. [46, 47]):

$$\Xi = \sum_i \exp\left(-\frac{E_i}{k_B T}\right). \tag{B.1}$$

Here $E_i$ is the energy level of state $i$ and summation is performed over all the states of the system.

Let us consider the magnetic defect system in the external field. Each defect consists of fermion spin pairs aligned into the cluster with a spin s=0, ±1. Taking into account exchange interaction between the spins inside any pair and neglecting interaction between the pairs (diluted magnetic system) one could get closed form expressions for the two limiting cases. Namely, for the case of all the spins pointing along the one preferential direction the partition function per defect is

$$\Xi = \exp\left(\frac{J + \mu_B H \cos(\theta)}{k_B T}\right) + \exp\left(\frac{J - \mu_B H \cos(\theta)}{k_B T}\right) + 2\exp\left(\frac{-J}{k_B T}\right). \tag{B.2a}$$

Here $J$ is the exchange energy, $H$ is an external magnetic field strength, $\mu_B$ is an elementary magnetic moment and the angle $\theta$ determines the direction of field with respect to the spins.

When we have the pairs able to rotate freely along the direction of the external field the partition function is

$$\Xi = 2\exp\left(\frac{J}{k_B T}\right)\sinh\left(\frac{\mu_B H}{k_B T}\right)\frac{k_B T}{\mu_B H} + 2\exp\left(\frac{-J}{k_B T}\right). \tag{B.2b}$$

Here we integrate over all the possible orientations of spin pairs. Then the free energy, magnetization and susceptibility $\chi$ of identical $N$ defects could be obtained as

$$F = -N k_B T \log(\Xi) \tag{B.3a}$$

$$M = -\frac{\partial F}{\partial H}, \qquad \chi = \frac{\partial M}{\partial H} = -\frac{\partial^2 F}{\partial H^2} \tag{B.3b}$$

Using the evident expressions (B.2) and (B.3), one could get the following magnetization dependences on the magnetic field:

$$M = \mu_B N \sinh\left(\frac{\mu_B H \cos(\theta)}{k_B T}\right) \bigg/ \left(\cosh\left(\frac{\mu_B H \cos(\theta)}{k_B T}\right) + \exp\left(\frac{-2J}{k_B T}\right)\right) \tag{B.4a}$$

for the spins pairs aligned at the angle θ with respect to the magnetic filed. While for the spin pairs able to rotate freely

$$M = \mu_B N \left(\left(\cosh\left(\frac{\mu_B H}{k_B T}\right) + \exp\left(\frac{-2J}{k_B T}\right)\right) \bigg/ \left(\sinh\left(\frac{\mu_B H}{k_B T}\right) + \exp\left(\frac{-2J}{k_B T}\right)\frac{\mu_B H}{k_B T}\right) - \frac{k_B T}{\mu_B H}\right) \tag{B.4b}$$

$$\chi = \mu_B N \left( \frac{k_B T}{\mu_B H^2} - \frac{1}{k_B T} \frac{1 + \exp\left(\frac{4J}{k_B T}\right) + \exp\left(\frac{2J}{k_B T}\right)\left(2\cosh\left(\frac{\mu_B H}{k_B T}\right) - \sinh\left(\frac{\mu_B H}{k_B T}\right)\frac{\mu_B H}{k_B T}\right)}{\left(\sinh\left(\frac{\mu_B H}{k_B T}\right)\exp\left(\frac{2J}{k_B T}\right) + \frac{\mu_B H}{k_B T}\right)^2} \right) \quad (B.4c)$$

For numerical simulations the exchange integral $J(T,R,N_{2D}) \approx \frac{J_m}{\varepsilon^2(T,R)} f\left(\frac{\varepsilon^{-1}(T,R)}{a_B} \cdot \sqrt{\frac{4}{\pi N_{2D}}}\right)$ depends on the temperature $T$, particle radius $R$ and defects concentration $N_{2D}$. The average distance $r$ between magnetic defects is calculated within percolation model as $r = \sqrt{\frac{4}{\pi N_{2D}}}$.

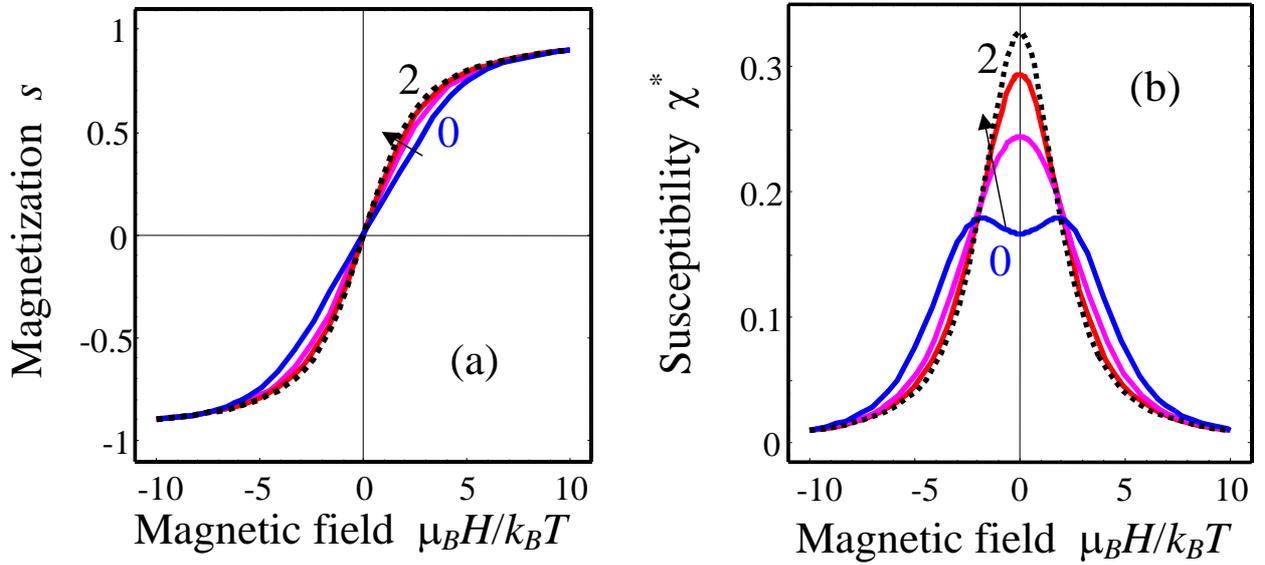

**Fig. B1.** Dependence of the magnetic order parameter $s = \frac{M}{\pi N_{2D} R^2 \mu_B}$ (a) and susceptibility $\chi^* = \chi \frac{k_B T}{\mu_B^2}$ (b) on the dimensionless magnetic field $\mu_B H/(k_B T)$ for different values of the dimensionless ratio $J/(k_B T) = 0, 0.5, 1, \geq 2$.

When is the defect concentration $N_{2D}$ and the particle radius $R$ high enough to use the mean field approach? Quantitatively this happens when
(1) the total amount of magnetic spins at the particle surface $N = \pi N_{2D} R^2$ is enough to introduce the continuous approximation, i.e. $\pi N_{2D} R^2 \gg 1$;
(2) the surface curvature should be small in comparison with the average distance between the surface defects $r$, i.e. $R \gg \sqrt{4/\pi N_{2D}}$.

Both conditions reduce to one strong inequality, $\pi N_{2D} R^2 >> 4$. The validity of this allows one to consider the effective Hamiltonian within Ising model:

$$\hat{H}_{eff} = -\frac{1}{2}\sum_{i,j} J_{ij} s_i s_j - \frac{1}{4}\sum_{i,j,k,m} J_{ijkm} s_i s_j s_k s_m - \mu_B \sum_i H_i s_i \qquad (B.5)$$

**H** is the magnetic field applied to the particle, $s_i$ is the defect spins. $J_{ij}$ and $J_{ijkm}$ are exchange energies of the two- and four-spins interactions respectively. With the help of Hamiltonian (B.5) in the molecular field approximation, one can obtain the following free energy for the system of spins [48, 49, 50]:

$$F = -\frac{1}{2}\sum_j J_{ij} \langle s_i \rangle \langle s_j \rangle - \frac{1}{4}\sum_{i,j,k,m} J_{ijkm} \langle s_i \rangle \langle s_j \rangle \langle s_k \rangle \langle s_m \rangle - \mu_B \sum_i H_i \langle s_i \rangle + \\ + k_B T \sum_i \left( \frac{1+\langle s_i \rangle}{2} \ln\left(\frac{1+\langle s_i \rangle}{2}\right) + \frac{1-\langle s_i \rangle}{2} \ln\left(\frac{1-\langle s_i \rangle}{2}\right) \right) \qquad (B.6)$$

Here $\langle s_i \rangle$ determines the thermally averaged spin "*i*" or the order parameter distribution across the system. The last sum represents the entropy of the system. Minimization of Eq. (B.5) leads to the following equation for the thermally averaged spins $\langle s_i \rangle$:

$$-\frac{1}{2}\sum_j J_{ij} \langle s_j \rangle - \frac{1}{4}\sum_{j,k,m} J_{ijkm} \langle s_j \rangle \langle s_k \rangle \langle s_m \rangle - \mu_B H_i + k_B T \operatorname{arctanh}\langle s_i \rangle = 0. \qquad (B.7)$$

Using the continuous approximation $\langle s_i \rangle \equiv s(\mathbf{r})$ and the relationship $s(\mathbf{r} \pm \mathbf{a}) = s(\mathbf{r}) \pm (\mathbf{a}\nabla) s(\mathbf{r}) + (\mathbf{a}\nabla)^2 s(\mathbf{r})/2 \pm \ldots$ (here **a** is the vector determining the position of the spin) and accounting for the spin temporal relaxation it is possible to derive the time-dependent equation for the spatial distribution of the order parameter, i. e. the fraction of coherently oriented spins $0 \leq s \leq 1$:

$$-\Gamma \frac{\partial}{\partial t} s - J s - J \delta \Delta s - J_{nl} s^3 - \mu_B H + k_B T \operatorname{arctanh}(s) = 0. \qquad (B.8)$$

Here $\Gamma$ is the relaxation coefficient, $J(T, R, N_{2D}) \approx \frac{J_m}{\varepsilon^2(T,R)} f\left( \frac{\varepsilon^{-1}(T,R)}{a_B} \cdot \sqrt{\frac{4}{\pi N_{2D}}} \right)$ is the effective exchange constant (the mean field acting on the each spin from its neighbors), $J_{nl}$ is the effective nonlinearity coefficient (unknown for incipient ferroelectrics, but typically a very small correction in comparison with the term $k_B T \operatorname{arctanh} s$). The gradient coefficient $\delta$ determines the correlation between the spins. Neglecting the gradient effects, the static Eq.(B.8) reduces to the known equation of state $s = \tanh\left( \frac{\mu_B H + J s + J_{nl} s^3}{k_B T} \right)$ for the equilibrium order parameter *s* [51].